\documentclass[unsortedaddress,aps,preprint,showpacs]{revtex4}
\usepackage{graphicx}
\usepackage{dcolumn}
\usepackage{amsmath}
\usepackage{amssymb}
\usepackage{amsfonts}
\usepackage{bm}
\usepackage{color}
\usepackage[normalem]{ulem}
\newcommand{\be}{\begin{equation}}
\newcommand{\ee}{\end{equation}}
\newcommand{\ba}{\begin{eqnarray}}
\newcommand{\ea}{\end{eqnarray}}

\begin{document}
\title{Freezing of soft-core bosons at zero temperature:\\a variational theory}
\author{Santi Prestipino$^1$\footnote{Corresponding author. Email: {\tt sprestipino@unime.it}}, Alessandro Sergi$^{1,2}$\footnote{Email: {\tt asergi@unime.it}}, and Ezio Bruno$^3$\footnote{Email: {\tt ebruno@unime.it}}}
\affiliation{$^1$Universit\`a degli Studi di Messina,\\Dipartimento di Scienze Matematiche e Informatiche, Scienze Fisiche e Scienze della Terra,\\viale F. Stagno d'Alcontres 31, 98166 Messina, Italy\\$^2$Institute of Systems Science, Durban University of Technology, P.\,O.\,Box 1334, Durban 4000, South Africa\\$^3$Universit\`a degli Studi di Messina,\\Dipartimento di Ingegneria,\\c.da Di Dio, 98166 S. Agata, Messina, Italy}
\date{\today}

\begin{abstract}
The properties of a macroscopic assembly of weakly-repulsive bosons at zero temperature are well described by Gross-Pitaevskii mean-field theory. According to this formalism the system exhibits a quantum transition from superfluid to cluster supersolid as a function of pressure. We develop a thermodynamically rigorous treatment of the different phases of the system by adopting a variational formulation of the condensate wave function --- represented as a sum of Gaussians --- that is amenable to exact manipulations. Not only is this description quantitatively accurate, but it is also capable to predict the order (and sometimes even the location) of the transition. We consider a number of crystal structures in two and three dimensions and determine the phase diagram. Depending on the lattice, the transition from fluid to solid can be first-order or continuous, a lower coordination entailing a milder transition. In two dimensions, crystallization would occur at the same pressure on three distinct lattices (square, honeycomb, and stripes), all providing metastable phases with respect to the triangular crystal. A similar scenario holds in three dimensions, where the simple-cubic and diamond crystals also share a common melting point; however, the stable crystal at low pressure is typically fcc. Upon compression and depending on the shape of the potential, the fcc crystal may transform into hcp. We conclude by sketching a theory of the solid-fluid interface and of quantum nucleation of the solid from the fluid.
\end{abstract}

\pacs{64.70.D-, 67.85.Bc, 67.80.K-}
\maketitle
\section{Introduction}

The experimental realization of Bose-Einstein condensation in trapped gases of alkali atoms in the nineties~\cite{Anderson,Davis}, made eventually possible by the development of novel (laser and evaporative) cooling techniques, has boosted a lot of theoretical and experimental activities on ultracold quantum systems (see, e.g., \cite{Pitaevskii&Stringari}). Generally speaking, these systems provide an opportunity to study quantum many-body effects under controlled conditions, even beyond the contact-interaction approximation assumed in the Bogoliubov theory~\cite{Bogoliubov}. In the weak-interaction limit, an effective approach to the physics of ultracold atoms is the simple mean-field theory, as formulated in terms of the Gross-Pitaevskii equation~\cite{Gross1,Pitaevskii,Gross2}. 

Interestingly, many quantum systems undergo phase transitions near zero temperature ($T=0$). Such transitions take place in many-body systems with competing ground states; they are driven by a non-thermal control parameter, such as pressure, magnetic field, or chemical composition. At the transition point, order is destroyed solely by quantum fluctuations. A quantum transition is continuous when the ground state of the system changes continuously across the transition point; otherwise, the transition is first-order. For instance, dipolar bosons confined in a one-dimensional optical lattice exhibit various phases as the strength of interaction increases, going from superfluid to a crystal-like state~\cite{Astrakharchik,Deuretzbacher,Zoellner}.

A paradigmatic example of quantum transition is the crystallization of softly-repulsive bosons at $T=0$~\cite{Pomeau,Josserand,Sepulveda,Cinti1,Kunimi,Cinti2,Ancilotto}. Experimental candidates for this transition are ultracold gases of atoms dressed with Rydberg states, which are highly-excited electronic states (see, e.g., \cite{Balewski}). The effective atom-atom interaction is a bounded pair repulsion, having an essentially flat core of micrometric radius and a positive van der Waals tail~\cite{Henkel1,Henkel2}. In classical terms, an interaction that is everywhere finite can stabilize cluster crystals at low temperature and high density~\cite{Likos,Zhang1,Zhang2,Prestipino1,Prestipino2}, based on purely energetic considerations~\cite{Mladek1}: for example, when repulsion is ``fatter'' than Gaussian, it is more convenient to form isolated blobs of particles than having them distributed homogeneously in space. Such an arrangement ensures a large mobility to atoms, which can freely hop from one site to another~\cite{Moreno}. Cluster-crystal order also occurs in weakly-repulsive bosons at high pressure, with the additional bonus of supersolid behavior ({\it i.e.}, crystalline order coexisting with superfluid behavior) near the melting point~\cite{Saccani1,Saccani2,Boninsegni,Macri1,Macri2}. 

Focusing on the penetrable-sphere model (PSM)~\cite{Marquest,Santos} as a prototype of bounded repulsion, we here provide a thorough variational study of the zero-temperature phase diagram of a thermodynamic system of identical bosons in two and three dimensions, thus completing a work initiated in Ref.\,\cite{Kunimi,Ancilotto}. Following an earlier proposal made by Tarazona~\cite{Tarazona} in the different context of classical density-functional theory, we assume a specific parametric form of the condensate wave function from the outset, first verifying that it indeed reproduces the optimal single-particle wave function and energy very accurately. The use of this variational state leads to a number of simplifications in the energy functional which make the theory much more manageable numerically, opening up to the possibility of working out the ground-state phase diagram of soft-core bosons in relatively small time. By considering a wide spectrum of possible lattices, we identify stable and metastable crystalline phases and fully characterize their melting transition. Moreover, we show that all these crystals are supersolid, {\it i.e.}, they exhibit non-classical rotational inertia. Finally, we present a preliminary discussion about the structure of the solid-fluid interface and of nucleation of the solid from the fluid.

The outline of the paper is as follows. In Sec.\,II we introduce the model and the variational theory employed to study its thermodynamics. We also outline the method used to analyze the transition behavior. In Sec.\,III we first assess the quality of our theory compared to the theory in Ref.\,\cite{Kunimi}; then, we present our results. Section IV is devoted to a mean-field description of the solid-fluid interface and of the ensuing theory of quantum nucleation. Concluding remarks are offered in Sec.\,V.

\section{Model and theory}
\setcounter{equation}{0}
\renewcommand{\theequation}{2.\arabic{equation}}

We consider a macroscopic number $N$ of point-like bosons of mass $m$, interacting through a {\em bounded} potential $u$, even function of its argument (an example is the PSM interaction, $u({\bf x})=\epsilon\Theta(\sigma-|{\bf x}|)$, where $\Theta$ is the Heaviside step function and $\epsilon,\sigma>0$). The system Hamiltonian reads:
\be
H=\sum_{i=1}^N\frac{p_i^2}{2m}+\sum_{i<j}u({\bf x}_i-{\bf x}_j)\,.
\label{eq2-1}
\ee
In the mean-field (Hartree) approximation, which applies for $u$ of sufficiently weak strength, the system ground state is represented as a perfect condensate:
\be
\Psi({\bf x}_1,\ldots,{\bf x}_N)=\prod_{i=1}^N\psi({\bf x}_i)
\label{eq2-2}
\ee
with
\be
\int_V{\rm d}^dx\,|\psi({\bf x})|^2=1\,,
\label{eq2-3}
\ee
where $d$ is the space dimensionality and $V={\cal O}(N)$. The single-particle state $\psi$ is chosen such that the expectation value of $H$ in the state $\Psi$ be as low as possible, which leads to (see, e.g., \cite{Rogel-Salazar}):
\be
-\frac{\hbar^2}{2m}\nabla^2\psi({\bf x})+N\int{\rm d}^dx'\,|\psi({\bf x}')|^2u({\bf x}-{\bf x}')\psi({\bf x})=\mu\psi({\bf x})\,.
\label{eq2-4}
\ee
The quantity $\mu$ in Eq.\,(\ref{eq2-4}) is the Lagrange multiplier enforcing the condition $\left\langle\Psi|\Psi\right\rangle=1$ (equivalent to Eq.\,(\ref{eq2-3})). In the quantum-gas literature, the above equation is known as the (time-independent) Gross-Pitaevskii (GP) equation. Clearly, Eq.\,(\ref{eq2-4}) is only a necessary condition; among all solutions, the physical one has the least possible energy.

Equation (\ref{eq2-4}) has always a spatially homogeneous solution. However, under appropriate conditions, crystalline order may develop. Hence, it is natural to use a plane waves expansion for the single-particle wave function:
\be
\psi({\bf x})=\frac{1}{\sqrt{V}}\sum_{\bf G}c_{\bf G}e^{i{\bf G}\cdot{\bf x}}\,,
\label{eq2-5}
\ee
where the ${\bf G}$'s are reciprocal-lattice vectors and $\sum_{\bf G}|c_{\bf G}|^2=1$; $V$ is the system volume and periodic conditions hold. This leads to rewrite the GP equation as~\cite{Kunimi}:
\be
\left(\frac{\hbar^2K^2}{2m}+\rho\widetilde{u}(0)\right)c_{\bf K}+\rho\sum_{{\bf G}\ne 0}\widetilde{u}(G)S_{\bf G}c_{{\bf K}+{\bf G}}=\mu c_{\bf K}\,,
\label{eq2-6}
\ee
where $\rho=N/V$ is the number density, $S_{\bf G}=\sum_{{\bf G}'}c_{{\bf G}'}c_{{\bf G}'+{\bf G}}^*$, and $\widetilde{u}(k)$ is the real-valued Fourier transform of $u$. The fluid phase, corresponding to $c_{\bf G}=\delta_{{\bf G},0}$, is a special solution to Eq.\,(\ref{eq2-6}) with $\mu=\rho\widetilde{u}(0)$.

A different but equivalent perspective is to view the Fourier coefficients $c_{\bf G}$, as well as the lattice constant $a$, as parameters to be optimized. Using the variational method, the best solution of type (\ref{eq2-5}) should minimize the average energy per particle:
\be
{\cal E}([c],a;\rho)=\frac{\hbar^2}{2m}\sum_{\bf G}G^2|c_{\bf G}|^2+\frac{\rho}{2}\sum_{{\bf G}_1,{\bf G}_2,{\bf G}_3}\widetilde{u}(G_1)c_{{\bf G}_1+{\bf G}_2}^*c_{{\bf G}_1+{\bf G}_3}c_{{\bf G}_2}c_{{\bf G}_3}^*\,,
\label{eq2-7}
\ee
{\it i.e.}, the sum of zero-point kinetic energy and potential energy. By requiring the derivative of ${\cal E}([c],a;\rho)-\mu\sum_{\bf G}|c_{\bf G}|^2$ with respect to $c_{\bf G}^*$ to be zero, we re-obtain Eqs.\,(\ref{eq2-6}). The way to solve these equations for a fixed $a$ is by iteration: at each step of the procedure, $S_{\bf G}$ is first estimated from the $c_{\bf G}$ coefficients drawn from the previous step; the resulting linear system is then solved, determining eigenvalues $\mu_n$ and normalized eigenvectors. Finally, the string of coefficients is updated to the eigenvector with the minimum ${\cal E}$ value.

Kunimi and Kato have solved Eq.\,(\ref{eq2-6}) for PSM bosons in two dimensions (2D)~\cite{Kunimi}, showing that for sufficiently high density the ground state is a triangular crystal (we shall later confirm and further extend their result by a rigorous thermodynamic analysis, see the end of this Section). Macr\`i {\em et al.}~\cite{Macri1} have tested mean-field (MF) results by Monte Carlo simulation, proving that the condensate is indeed almost perfect in the fluid region and that the exact freezing point lies extremely close to the theoretical estimate. However, if we wish to perform a systematic study of the phases of the PSM and systems alike in three dimensions, the effort of solving Eq.\,(\ref{eq2-6}) or to perform accurate simulations would be much greater. That is why we make an ansatz on the shape of $\psi$, described as a sum of Gaussians centered at the lattice sites, which is of no consequence for the overall picture since --- as we shall verify --- the results obtained are close to those of unconstrained MF theory.

We decide to represent the self-organized, quantum single-particle state by the real-valued wave function
\be
\psi({\bf x})=C_\alpha\frac{1}{\sqrt{V}}\sum_{\bf R}e^{-\alpha\left(\frac{{\bf x}-{\bf R}}{a}\right)^2}\,,
\label{eq2-8}
\ee
where $C_\alpha$ is a normalization constant and the ${\bf R}$'s are direct-lattice vectors. Two variational parameters appear in (\ref{eq2-8}), {\it i.e.}, $\alpha$ and $a$, respectively related to the width and periodicity of the Gaussians. We stress that $a$, to be interpreted hereafter as the nearest-neighbor distance, is an adjustable parameter as well, independent of the density, so as to grant the possibility to have cluster-crystal solutions (see the follow-up discussion at the end of this Section). When $\alpha\rightarrow 0$, the fluid phase is recovered.

Our first task is to normalize $\psi({\bf x})$, by requiring that Eq.\,(\ref{eq2-3}) is satisfied. Using the identity
\be
({\bf x}-{\bf R})^2+({\bf x}-{\bf R}')^2=2\left({\bf x}-\frac{{\bf R}+{\bf R}'}{2}\right)^2+\frac{({\bf R}-{\bf R}')^2}{2}\,,
\label{eq2-9}
\ee
$C_\alpha$ is easily found to be:
\be
C_\alpha=\sqrt{\frac{v_0}{a^d}}\left(\frac{2\alpha}{\pi I(\alpha)^{2/d}}\right)^{d/4}\,\,\,\,\,\,{\rm with}\,\,\,\,\,\,I(\alpha)=\sum_{\bf R}e^{-\frac{\alpha}{2a^2}R^2}
\label{eq2-10}
\ee
($v_0$ is the volume of the primitive cell, e.g., $v_0=(\sqrt{3}/2)a^2$ for the triangular lattice). On the other hand, $\psi({\bf x})$ can also be written as a Fourier series,
\be
\psi({\bf x})=\frac{1}{\sqrt{V}}\sum_{\bf G}\psi_{\bf G}e^{i{\bf G}\cdot{\bf x}}\,.
\label{eq2-11}
\ee
Denoting ${\cal C}$ a primitive cell, we find:
\be
\psi_{\bf G}=\sqrt{V}\frac{1}{v_0}\int_{\cal C}{\rm d}^dx\,e^{-i{\bf G}\cdot{\bf x}}\psi({\bf x})\equiv C_\alpha'e^{-\frac{G^2a^2}{4\alpha}}\,\,\,\,\,\,{\rm with}\,\,\,\,\,\,C_\alpha'=\sqrt{\frac{a^d}{v_0}}\left(\frac{2\pi}{\alpha I(\alpha)^{2/d}}\right)^{d/4}\,.
\label{eq2-12}
\ee
Finally, from the normalization condition $\sum_{\bf G}\psi_{\bf G}^2=1$ we derive a different expression for $I(\alpha)$:
\be
I(\alpha)=\frac{a^d}{v_0}\left(\frac{2\pi}{\alpha}\right)^{d/2}\sum_{\bf G}e^{-\frac{G^2a^2}{2\alpha}}\,,
\label{eq2-13}
\ee
which proves useful to develop a low-$\alpha$ expansion of the energy functional (see Appendix A).

The advantage of the Gaussian series (\ref{eq2-8}) over the more general expression (\ref{eq2-5}) is an analytical simplification of the energy functional, allowing a considerable speed up in the computations. Let us first consider the specific ({\it i.e.}, per unit particle) kinetic energy. Its general expression is
\be
{\cal E}_{\rm kin}=-\frac{\hbar^2}{2m}\int{\rm d}^dx\,\psi^*({\bf x})\nabla^2\psi({\bf x})=\frac{\hbar^2}{2m}\int{\rm d}^dx\,\nabla\psi^*({\bf x})\cdot\nabla\psi({\bf x})\,,
\label{eq2-14}
\ee
where the equality follows after observing that, even though $\psi$ and its gradient do not vanish at infinity, the integral over a cell of every partial derivative of a smooth periodic function is zero. For the function $\psi$ in Eq.\,(\ref{eq2-8}) it readily follows that
\be
\int{\rm d}^dx\left(\nabla\psi({\bf x})\right)^2=\frac{4\alpha^2C_\alpha^2}{Va^4}\sum_{{\bf R},{\bf R}'}e^{-\frac{\alpha}{2a^2}({\bf R}-{\bf R}')^2}\int{\rm d}^dx\,({\bf x}-{\bf R})\cdot({\bf x}-{\bf R}')e^{-\frac{2\alpha}{a^2}\left({\bf x}-\frac{{\bf R}+{\bf R}'}{2}\right)^2}\,.
\label{eq2-15}
\ee
The inner integral is solved by a change of variables, eventually arriving at:
\be
\int{\rm d}^dx\left(\nabla\psi({\bf x})\right)^2=\frac{4\alpha^2C_\alpha^2}{v_0a^4}\left(\frac{\pi a^2}{2\alpha}\right)^{d/2}\sum_{\bf R}\left(\frac{da^2}{4\alpha}-\frac{R^2}{4}\right)e^{-\frac{\alpha}{2a^2}R^2}\,.
\label{eq2-16}
\ee
Finally, using Eq.\,(\ref{eq2-13}) and its derivative, a closed-form expression for the kinetic energy is obtained:
\be
{\cal E}_{\rm kin}=e_0\frac{d}{2}\frac{\alpha\sigma^2}{a^2}\left(1+\frac{2}{d}\alpha\frac{I'(\alpha)}{I(\alpha)}\right)\,.
\label{eq2-17}
\ee
In this equation, $\sigma$ is a characteristic length of the potential (say, its range) whereas $e_0=\hbar^2/(m\sigma^2)$ is a natural energy unit. Using these units, we see from (\ref{eq2-7}) that the ground state is only controlled by the dimensionless quantity $\rho\sigma^d\epsilon/e_0$ (which we refer in the following as the ``density'') or, equivalently, by the value of $g\equiv\rho\widetilde{u}(0)/e_0$. 

As for the potential energy, it admits no concise form like (\ref{eq2-17}), but its expression can nevertheless be greatly simplified and reduced to the numerical evaluation of a few single, rapidly converging series. Indeed, replacing $c_{\bf G}$ with $\psi_{\bf G}$ in the second term of (\ref{eq2-7}) we get:
\be
{\cal E}_{\rm pot}=\frac{1}{2}\rho C_\alpha^{\prime 4}\sum_{{\bf G}_1}\widetilde{u}(G_1)e^{-\frac{G_1^2a^2}{2\alpha}}\sum_{{\bf G}_2}e^{-\frac{(G_2^2+{\bf G}_1\cdot{\bf G}_2)a^2}{2\alpha}}\sum_{{\bf G}_3}e^{-\frac{(G_3^2+{\bf G}_1\cdot{\bf G}_3)a^2}{2\alpha}}\,.
\label{eq2-18}
\ee
Each of the inner sums equals:
\be
\sum_{\bf G}e^{-\frac{(G^2+{\bf G}_1\cdot{\bf G})a^2}{2\alpha}}=e^{\frac{G_1^2}{8\alpha}}\sum_{\bf G}e^{-\frac{({\bf G}+{\bf G}_1/2)^2a^2}{2\alpha}}\,.
\label{eq2-19}
\ee
For the sake of clarity, now take $d=2$. Writing ${\bf G}_1$ as an integer combination of reciprocal-lattice basis vectors, {\it i.e.}, ${\bf G}_1=p{\bf b}_1+q{\bf b}_2$, the sum in the r.h.s. of Eq.\,(\ref{eq2-19}) can at most take 4 ($=2^d$) distinct values, according to whether $p,q$ are even or odd:
\be
\sum_{\bf G}e^{-\frac{({\bf G}+{\bf G}_1/2)^2a^2}{2\alpha}}=\left\{
\begin{array}{rl}
\sum_{\bf G}e^{-\frac{G^2a^2}{2\alpha}}\,, & p\,\,{\rm and}\,\,q\,\,{\rm both}\,\,{\rm even} \\
\sum_{\bf G}e^{-\frac{({\bf G}+{\bf b}_1/2)^2a^2}{2\alpha}}\,, & p\,\,{\rm odd\,\,and}\,\,q\,\,{\rm even} \\
\sum_{\bf G}e^{-\frac{({\bf G}+{\bf b}_2/2)^2a^2}{2\alpha}}\,, & p\,\,{\rm even\,\,and}\,\,q\,\,{\rm odd} \\
\sum_{\bf G}e^{-\frac{({\bf G}+({\bf b}_1+{\bf b}_2)/2)^2a^2}{2\alpha}}\,, & p\,\,{\rm and}\,\,q\,\,{\rm both}\,\,{\rm odd} 
\end{array}
\right.\,.
\label{eq2-20}
\ee
Denoting $J_1(\alpha),J_2(\alpha),J_3(\alpha)$, and $J_4(\alpha)$ the four sums in Eq.\,(\ref{eq2-20}), the specific potential energy becomes (with obvious meaning of the symbols):
\ba
{\cal E}_{\rm pot}&=&\frac{1}{2}\rho C_\alpha^{\prime 4}\left\{J_1(\alpha)^2\sum_{{\bf G}}^{\rm (e,e)}\widetilde{u}(G)e^{-\frac{G^2a^2}{4\alpha}}+J_2(\alpha)^2\sum_{{\bf G}}^{\rm (o,e)}\widetilde{u}(G)e^{-\frac{G^2a^2}{4\alpha}}\right.
\nonumber \\
&+&\left.J_3(\alpha)^2\sum_{{\bf G}}^{\rm (e,o)}\widetilde{u}(G)e^{-\frac{G^2a^2}{4\alpha}}+J_4(\alpha)^2\sum_{{\bf G}}^{\rm (o,o)}\widetilde{u}(G)e^{-\frac{G^2a^2}{4\alpha}}\right\}\,.
\label{eq2-21}
\ea
Further simplifications may occur depending on the lattice. For example, while $J_2(\alpha)=J_3(\alpha)=J_4(\alpha)$ on the triangular lattice, $J_2(\alpha)=J_3(\alpha)\ne J_4(\alpha)$ on the square lattice. In the former case, the energy per particle reads in compact form:
{\small
\be
{\cal E}(\alpha,a;\rho)=e_0\frac{\alpha\sigma^2}{a^2}\left(1+\alpha\frac{I'(\alpha)}{I(\alpha)}\right)+\frac{1}{2}\rho\left\{\sum_{{\bf G}}^{\rm (e,e)}\widetilde{u}(G)e^{-\frac{G^2a^2}{4\alpha}}+\left(\frac{J(\alpha)}{I(\alpha)}\right)^2\sum_{{\bf G}}^{\neg{\rm (e,e)}}\widetilde{u}(G)e^{-\frac{G^2a^2}{4\alpha}}\right\}
\label{eq2-22}
\ee
}
with
\be
J(\alpha)=\frac{2\pi a^2}{\alpha v_0}\sum_{\bf G}e^{-\frac{({\bf G}+{\bf b}_1/2)^2a^2}{2\alpha}}\,.
\label{eq2-23}
\ee
For the triangular lattice, the reciprocal-lattice vectors are given by:
\be
{\bf b}_1=\frac{2\pi}{a}\left(1,-\frac{1}{\sqrt{3}}\right)\,\,\,\,\,\,{\rm and}\,\,\,\,\,\,{\bf b}_2=\frac{2\pi}{a}\left(0,\frac{2}{\sqrt{3}}\right)\,.
\label{eq2-24}
\ee
Numerical minimization of Eq.\,(\ref{eq2-21}) will give the optimal $\alpha$ and $a$. An expression similar to (\ref{eq2-21}) holds for any Bravais lattice.

The situation is somewhat harder for a non-Bravais lattice ({\it i.e.}, a Bravais lattice with a basis). An example is the honeycomb lattice: its reference lattice is triangular with lattice constant $c=\sqrt{3}a$, but every cell of volume $v_0=(\sqrt{3}/2)c^2$ contains two particles, whose positions within the cell are described by, say, ${\bf e}_1=(0,0)$ and ${\bf e}_2=(0,a)$. The variational wave function now reads:
\be
\psi({\bf x})=C_\alpha\frac{1}{\sqrt{V}}\sum_{{\bf R},{\bf e}}e^{-\alpha\left(\frac{{\bf x}-{\bf R}-{\bf e}}{a}\right)^2}\,,
\label{eq2-25}
\ee
where $C_\alpha$ is still given by Eq.\,(\ref{eq2-10}), but $I(\alpha)$ is different:
\be
I(\alpha)=\sum_{\bf R}\left[2e^{-\frac{\alpha}{2a^2}{\bf R}^2}+e^{-\frac{\alpha}{2a^2}({\bf R}+{\bf e}_1-{\bf e}_2)^2}+e^{-\frac{\alpha}{2a^2}({\bf R}+{\bf e}_2-{\bf e}_1)^2}\right]\,.
\label{eq2-26}
\ee
The Fourier coefficients of $\psi({\bf x})$ are now written as:
\be
\psi_{\bf G}=C_\alpha'\left(e^{-i{\bf G}\cdot{\bf e}_1}+e^{-i{\bf G}\cdot{\bf e}_2}\right)e^{-\frac{G^2a^2}{4\alpha}}\,,
\label{eq2-27}
\ee
with the same $C_\alpha'$ as in Eq.\,(\ref{eq2-12}). Imposing normalization in the Fourier representation, an alternate $I(\alpha)$ expression follows:
\be
I(\alpha)=4\left(\frac{2\pi}{\alpha}\right)^{d/2}\frac{a^d}{v_0}\sum_{\bf G}\cos^2\left(\frac{{\bf G}\cdot({\bf e}_2-{\bf e}_1)}{2}\right)e^{-\frac{G^2a^2}{2\alpha}}\,.
\label{eq2-28}
\ee
Finally, the energy per particle is given by
\be
{\cal E}=\frac{1}{2}e_0\sum_{\bf G}(G\sigma)^2|\psi_{\bf G}|^2+\frac{1}{2}\rho\sum_{{\bf G}_1,{\bf G}_2,{\bf G}_3}\widetilde{u}(G_1)\psi_{{\bf G}_1+{\bf G}_2}^*\psi_{{\bf G}_1+{\bf G}_3}\psi_{{\bf G}_2}\psi_{{\bf G}_3}^*\,,
\label{eq2-29}
\ee
where, using $\psi_{\bf G}=u_{\bf G}+iv_{\bf G}$:
\ba
\sum_{{\bf G}_2,{\bf G}_3}\psi_{{\bf G}_1+{\bf G}_2}^*\psi_{{\bf G}_1+{\bf G}_3}\psi_{{\bf G}_2}\psi_{{\bf G}_3}^*&=&\sum_{{\bf G}_2,{\bf G}_3}\left[(u_{{\bf G}_1+{\bf G}_2}u_{{\bf G}_1+{\bf G}_3}+v_{{\bf G}_1+{\bf G}_2}v_{{\bf G}_1+{\bf G}_3})(u_{{\bf G}_2}u_{{\bf G}_3}+v_{{\bf G}_2}v_{{\bf G}_3})\right.
\nonumber \\
&+&\left.(u_{{\bf G}_1+{\bf G}_2}v_{{\bf G}_1+{\bf G}_3}-v_{{\bf G}_1+{\bf G}_2}u_{{\bf G}_1+{\bf G}_3})(u_{{\bf G}_2}v_{{\bf G}_3}-v_{{\bf G}_2}u_{{\bf G}_3})\right]\,.
\nonumber \\
\label{eq2-30}
\ea

%
%
\begin{figure}
\begin{center}
\includegraphics[width=15cm]{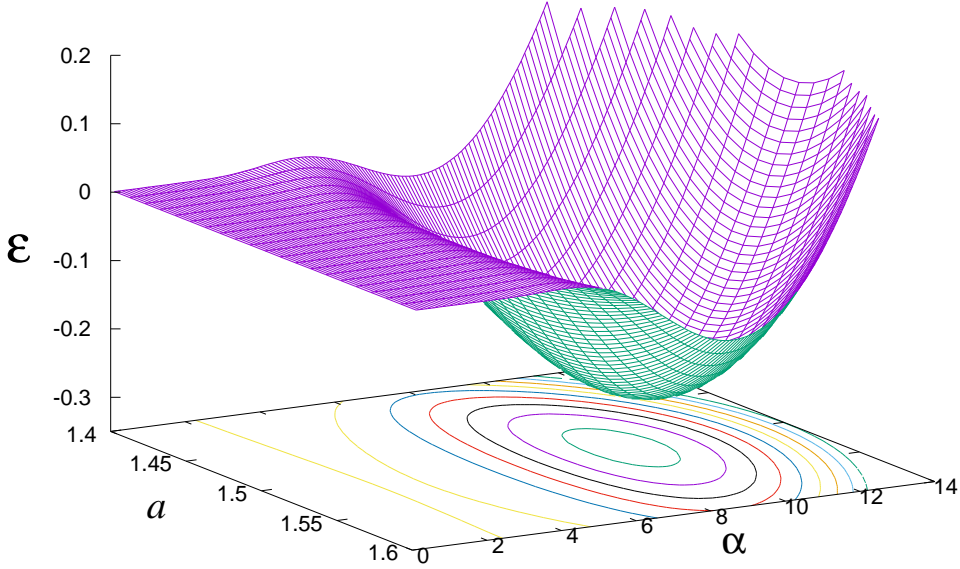}
\caption{PSM bosons on the triangular lattice at $T=0$: typical surface plot of the energy functional (\ref{eq2-22}) at high density (note that the zero of energy has here been shifted to $\rho\widetilde{u}(0)/2$). For this example, which refers to $\rho=14$ (units of $e_0\epsilon^{-1}\sigma^{-d}$), the absolute minimum falls at $\alpha=9.50204$ and $a=1.504540$ (units of $\sigma$). The location of the minimum energy ($-0.2716735\ldots$) is separated by a low barrier from the ``fluid'' minimum at $\alpha=0$ (to make this barrier visible, contour lines have been plotted every 0.05 starting from $-0.2499$).}
\label{fig1}
\end{center}
\end{figure}

Before closing this Section, we discuss how to extract thermodynamic properties from raw energy data. Once best parameters ($\overline{\alpha}$ and $\overline{a}$) have been computed for each density, the internal energy per particle is given by $e(\rho)={\cal E}(\overline{\alpha}(\rho),\overline{a}(\rho);\rho)$ (there is a different energy branch for each crystal, while $e(\rho)=(\widetilde{u}(0)/2)\rho$ for the fluid). Typically, at low density the deepest minimum of ${\cal E}$ as a function of $\alpha$ occurs at $\alpha=0$. Upon increasing $\rho$, and provided that crystallization is first-order, a secondary minimum first appears at a positive $\alpha$ value, which then becomes the absolute minimum at a still larger density (see Fig.\,1). However, if the pressure $P$ is fixed the stable phase must minimize the {\em generalized enthalpy}, $\widetilde{h}(\rho;T=0,P)=e(\rho)+P/\rho$ (per unit particle). The minimum $\widetilde{h}$ is the enthalpy $h(P)$ at $T=0$, while the abscissa $\rho_{\rm eq}(P)$ of the minimum is the equilibrium density. Alternatively, we can resort to a graphical construction: for each possible phase, we plot $e$ as a function of the specific volume $v=\rho^{-1}$; the slope of the tangent line at $v$ is $-P(v)$. For a given lattice, the transition occurs where the fluid and crystal energy branches have a common tangent, and the coexistence volumes are the abscissae of the contact points. Finally, the chemical potential at $P$ is $\mu=e(\rho_{\rm eq})+P/\rho_{\rm eq}$, which is nothing but the intercept on the energy axis of the tangent at $(\rho_{\rm eq}^{-1},e(\rho_{\rm eq}))$. In formal terms, the full equilibrium energy curve coincides with the boundary of the convex hull of all the individual $e$ vs. $v$ curves.

We stress that in a crystal of soft-core particles the number $N_c$ of cells may not be equal to $N$. Indeed, the {\em classical} PSM interaction is known for stabilizing cluster crystals at low temperature~\cite{Zhang2}. The same will also occur, based on the argument in \cite{Likos}, for smoothed-step interactions like the softened van der Waals (SVDW) repulsion, $u(r)=\epsilon/[1+(r/\sigma)^6]$, and the sequence of generalized-exponential-model (GEM) potentials, $u_n(r)=\epsilon\exp\{-(r/\sigma)^n\}$, for $n>2$~\cite{Mladek2}. The Gaussian repulsion ($n=2$) is a marginal case: despite there is no evidence of a cluster crystal in two or three dimensions, clear hints of clusterization are detected in one dimension~\cite{Speranza}. The quantum counterparts of the PSM and SVDW interactions have been studied by Monte Carlo simulation in Refs.\,\cite{Saccani1,Cinti2,Cinti3}, and cluster crystals have been found. In a mean-field setting, the criterion for clusterization is simply stated as follows. Denoting $N_c$ the number of lattice cells, the number of particles per cell is on average:
\be
\frac{N}{N_c}=\frac{N}{V}\frac{V}{N_c}=\rho v_0\,.
\label{eq2-31}
\ee
Therefore, if in equilibrium $\rho v_0>1$ the crystalline phase is actually a cluster crystal.

\section{Results}
\setcounter{equation}{0}
\renewcommand{\theequation}{3.\arabic{equation}}

We first present results for PSM bosons in two dimensions. By numerically solving the GP equation, Kunimi and Kato have concluded that the fluid coexists at $T=0$ with a triangular crystal in the interval $38.44\le g\le 40.98$~\cite{Kunimi}. Besides confirming this result with our approach, we shall provide data for other metastable 2D crystals, showing that crystallization on non-triangular lattices would instead be continuous.

We first solve Eq.\,(\ref{eq2-6}) on the triangular lattice (${\bf G}=p{\bf b}_1+q{\bf b}_2$, with ${\bf b}_1$ and ${\bf b}_2$ defined at Eq.\,(\ref{eq2-24})). For fixed $\rho$ and $a$, we truncate the system of equations by assuming that $c_{\bf G}=0$ for $|p|,|q|>5$ (we have checked that nothing changes if this threshold were rather 10). Then, diagonalization of the resulting $121\times 121$ Hermitian matrix of coefficients is cyclically performed within the iterative procedure described in Section II, until self-consistency is attained. At this point, we verify that
\be
\mu=2{\cal E}-\frac{1}{2}e_0\sum_{\bf G}(G\sigma)^2|c_{\bf G}|^2\,,
\label{eq3-1}
\ee
as expected. Finally, $a$ is optimized until its value is determined to five decimal places. Next, for the same lattice we solve the variational theory, searching for the minimum of (\ref{eq2-22}) on a grid of $(\alpha,a)$ values covering the region where the absolute minimum of ${\cal E}$ lies. The spacing of the grid is progressively reduced around the minimum, until its location is determined to $10^{-6}$ precision. The whole procedure is then repeated for the square lattice.

%
%
\begin{figure}
\begin{center}
\includegraphics[width=16cm]{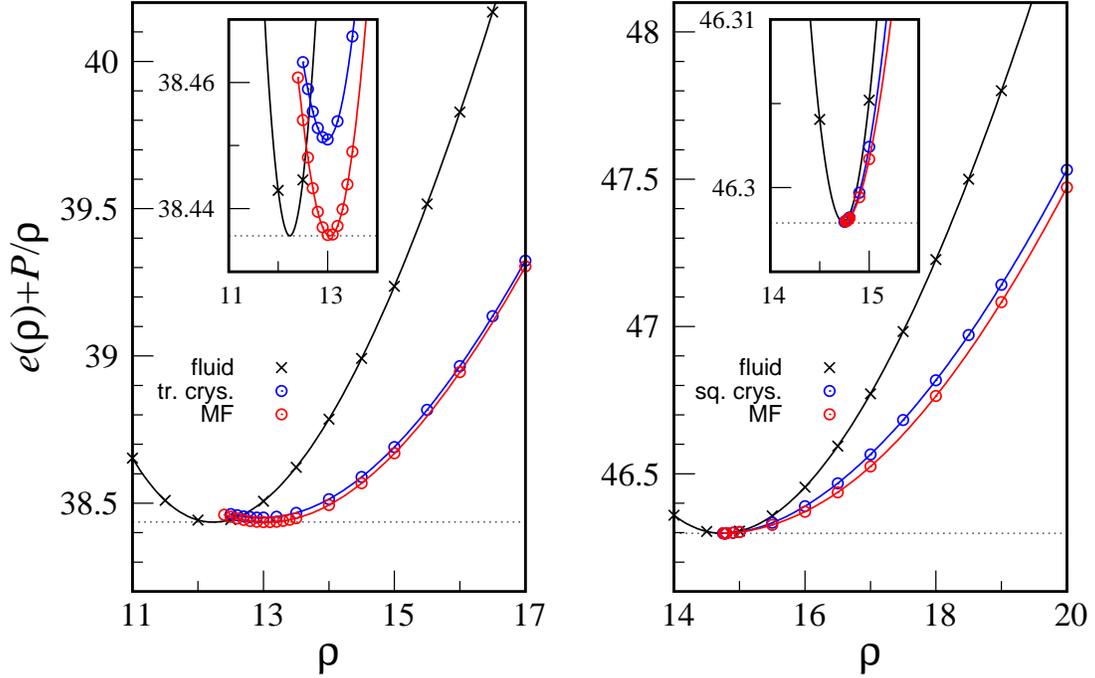}
\caption{PSM bosons in two dimensions at $T=0$: generalized enthalpy for two pressures, $P=235.12$ (left) and $P=341.1488$ (right). At these pressures, freezing in MF theory occurs into a triangular crystal and into a square crystal, respectively (in all figures, $e$ and $P$ values are given in units of $e_0$ and $e_0^2\epsilon^{-1}\sigma^{-d}$, respectively). Besides fluid data (black crosses), we report crystal data from MF theory (red dots) and Gaussian variational theory (blue dots). In the latter theory the transition to the triangular crystal occurs at $P=238.24$ (1.3\% higher than the MF estimate), whereas on the square lattice the transition pressure is the same for the two theories. In the insets, a magnification of the transition region is shown. Each horizontal dotted line marks the value of the chemical potential at the transition ($38.436$, left; $46.2979$, right).}
\label{fig2}
\end{center}
\end{figure}

%
%
\begin{figure}
\begin{center}
\includegraphics[width=16cm]{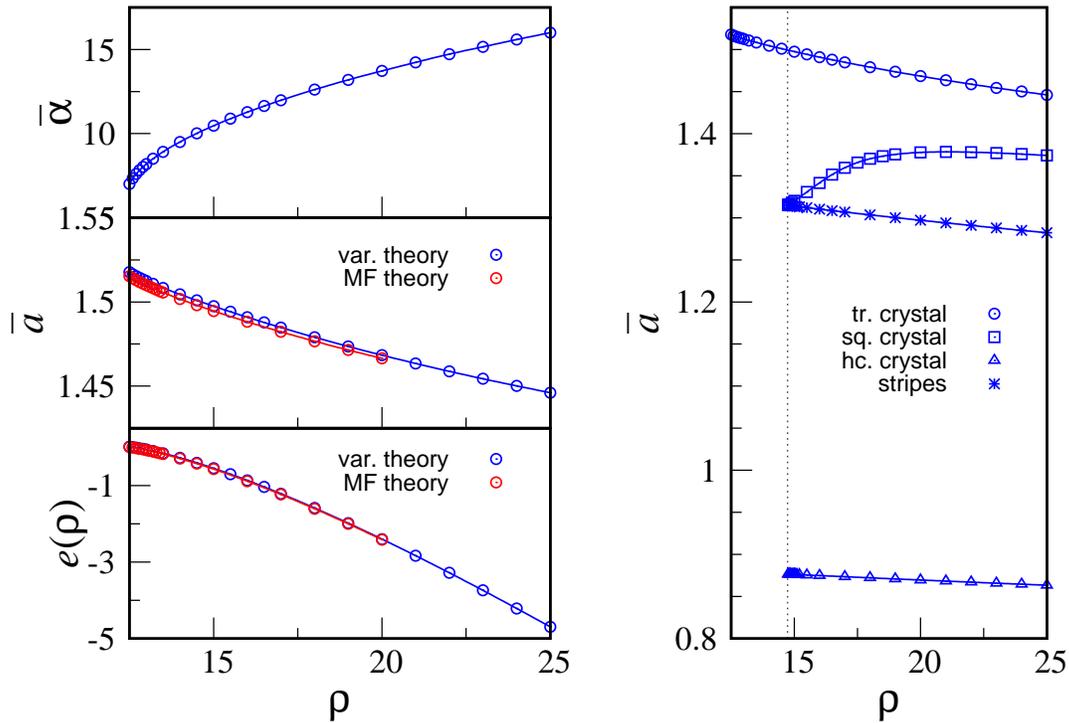}
\caption{PSM bosons in two dimensions at $T=0$. Left: optimal parameters for the triangular crystal as a function of the density $\rho$ according to MF theory (red dots) and Gaussian variational theory (blue dots). Right: variational-theory lattice constant $\overline{a}$, plotted as a function of $\rho$, for various 2D crystals. The vertical dotted line marks continuous freezing ($\rho=14.73710\ldots$, see Appendix A).}
\label{fig3}
\end{center}
\end{figure}

%
%
\begin{figure}
\begin{center}
\includegraphics[width=16cm]{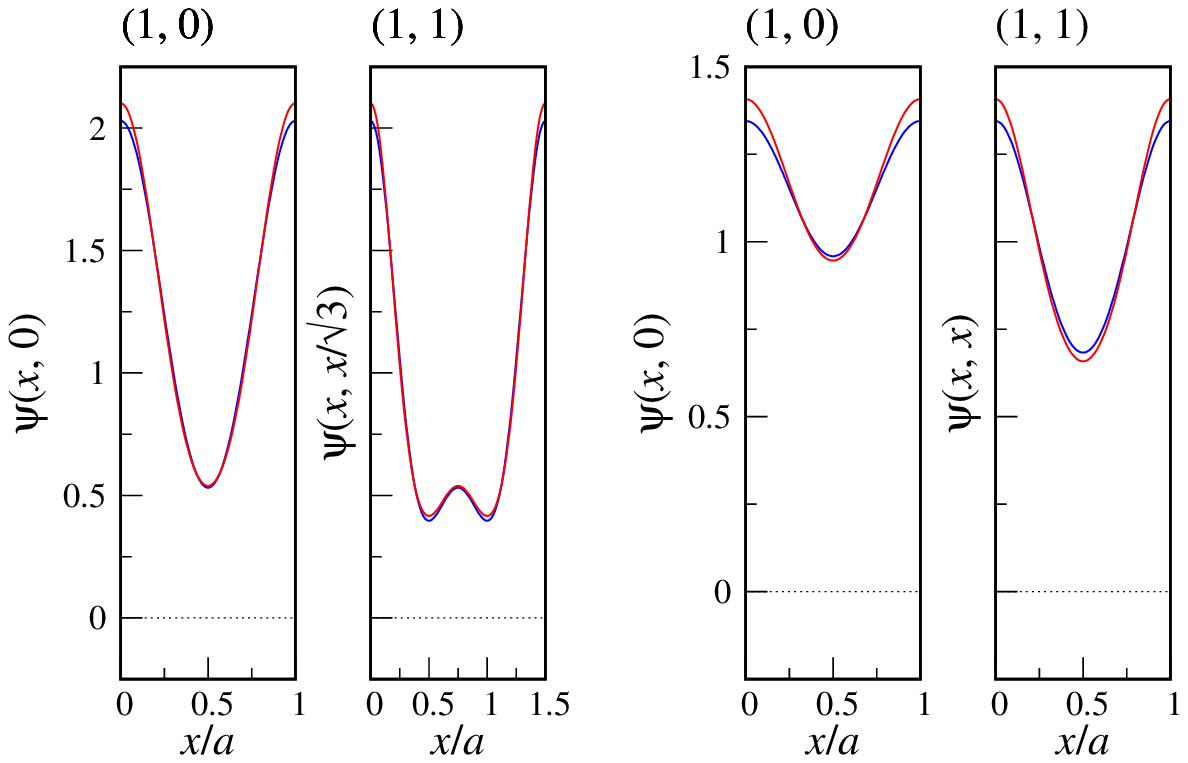}
\caption{PSM bosons in two dimensions at $T=0$: we compare the crystalline ground state $\psi$ in MF theory (red) with the optimal variational state $\psi$ (blue), along two space directions ($(1,0)$ and $(1,1)$) and on two lattices (triangular lattice: $\rho=13$, left panels; and square lattice: $\rho=15$, right panels).}
\label{fig4}
\end{center}
\end{figure}

The data in Figs.\,2 and 3 clearly document that the two theories give largely similar indications for the transition properties of PSM bosons at zero temperature. Looking at Fig.\,2, we see that freezing is first-order on the triangular lattice (left panel), while it seems to be continuous on the square lattice (right panel). As for the latter, we prove in Appendix A.1 that the transition to a square crystal indeed occurs continuously within Gaussian variational theory, by all evidence at the same density/pressure indicated by MF theory. For the case of the triangular crystal we report in Fig.\,3 left panel the optimal values of the variational parameters $\alpha$ and $a$. Again, $\overline{a}$ and $e(\rho)$ have nearly identical values at all densities in the two theories. Coexisting densities are $\rho_{\rm F0}=12.234$ and $\rho_{\rm S0}=13.045$ in MF theory (fully consistent with the $g$ thresholds reported in Ref.\,\cite{Kunimi}), while $\rho_{\rm F0}=12.315$ and $\rho_{\rm S0}=13.131$ (both 0.7\% higher) in variational theory. Finally, we show in Fig.\,4 a comparison between the crystalline ground states in the two theories close to melting. For both types of crystal, the wave functions along two distinct high-symmetry directions are, to a large extent, similar.

%
%
\begin{figure}
\begin{center}
\includegraphics[width=16cm]{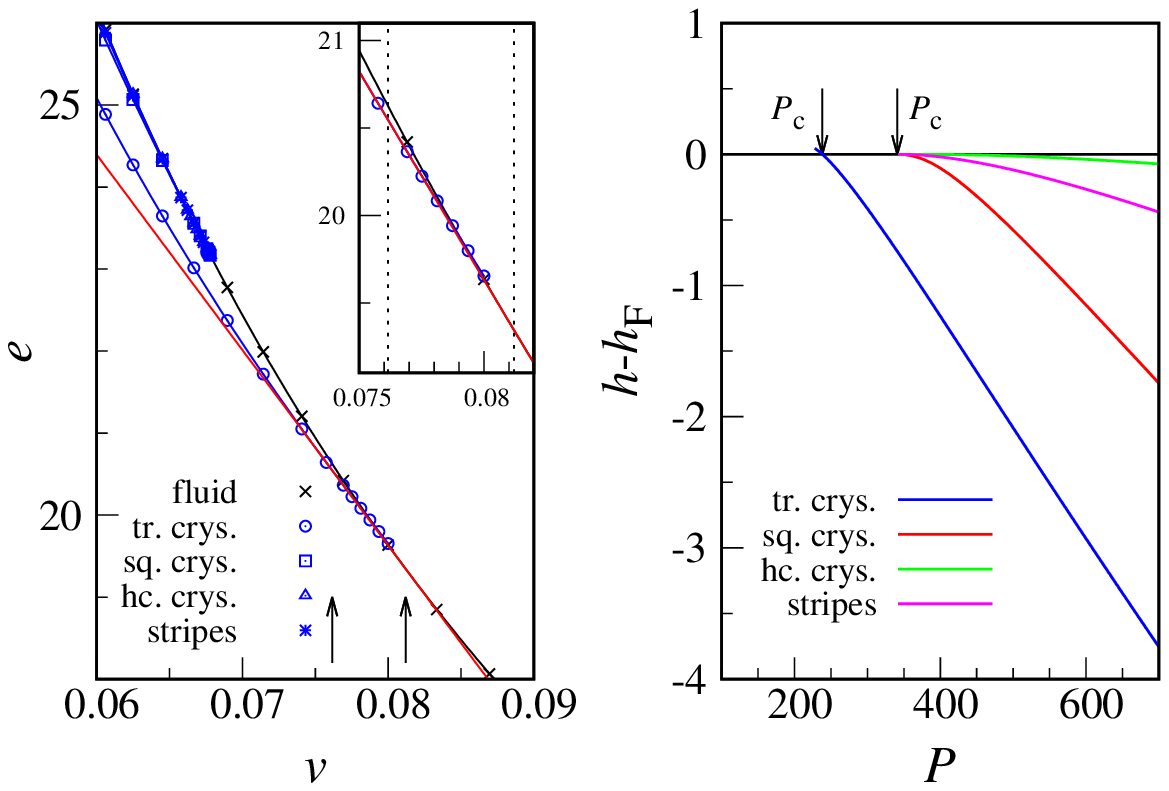}
\caption{PSM bosons in two dimensions at $T=0$, variational-theory results. Left: $e$ vs. $v$ for all the phases examined (see legend). The red straight line is the common tangent to the fluid and triangular-crystal branches. The arrows mark the coexisting volumes. The inset shows a magnification of the transition region (the coexisting volumes are now signaled by two dotted lines). Right: enthalpy $h(P)$ for each solid phase, using the fluid phase as reference (black). The arrows mark the location of the transition into the triangular crystal (left) and into the square crystal (right). The transition pressure for the honeycomb and the striped crystal is the same as for the square crystal.}
\label{fig5}
\end{center}
\end{figure}

Then, we have considered other crystals, a honeycomb crystal and a striped crystal (periodic in one direction only), to see what transition pressure would result in these cases. Like the square crystal, also these crystals melt continuously. More importantly, the melting pressure is apparently the same as for the square crystal. Indeed, we rigorously prove in Appendices A.2 and A.3 that the transition point is exactly the same for the three crystals, at least within Gaussian variational theory. This evidence is surprising: not only the nature of the transition is the same for the three lattices but also its location is universal (we shall come back to this later). In the right panel of Fig.\,3 we compare the values of $\overline{a}$ in the various crystals. As it might be expected, $\overline{a}$ typically decreases with increasing density, only the square crystal makes exception to this rule at moderate densities, signaling an anomalous behavior of the mean site occupancy.

%
%
\begin{table}
\caption{Soft-core bosons at $T=0$: location of the freezing transition (the error is of one unit on the last decimal place; in one case only the datum refers to a solid-solid transition). Where not specified, results are from Gaussian variational theory.}
\begin{center}
\begin{tabular}{ccccc}
\hline\hline
model & crystal & $P_c$ & $\mu_c$ & order\\
\hline
PSM & tr. (MF) & 235.12 & 38.436 & 1st\\
PSM & tr. & 238.24 & 38.690 & 1st\\
PSM & sq. (MF) & 341.1488 & 46.2979 & 2nd\\
PSM & sq. & 341.1488 & 46.2979 & 2nd\\
PSM & hc. & 341.1488 & 46.2979 & 2nd\\
PSM & stripes & 341.1488 & 46.2979 & 2nd\\
\hline
GEM-10 & tr. & 421.22 & 47.987 & 1st\\
GEM-4 & tr. & 942.89 & 66.516 & 1st\\
SVDW & tr. & 541.20 & 53.306 & 1st\\
\hline\hline
PSM & fcc & 427.89 & 59.872 & 1st\\
PSM & bcc & 430.13 & 60.029 & 1st\\
PSM & fcc\,$\rightarrow$\,hcp & 510.5 & 64.55 & 1st\\
PSM & sh & 705.80 & 76.895 & 1st\\
PSM & sc & 987.4772 & 90.9543 & 2nd\\
PSM & diam. & 987.4772 & 90.9543 & 2nd\\
\hline
SVDW & fcc & 1013.65 & 81.667 & 1st\\
\hline\hline
\end{tabular}
\end{center}
\end{table}

To establish which phase is stable at a given pressure there is no other way but to try all the many possibilities, compute the energy as a function of density for each, and finally select the one with the lowest enthalpy. We show the outcome in Fig.\,5: in the left panel the energy of each phase is plotted as a function of volume; in the right panel, the enthalpies of the various phases are compared with each other. As expected, the triangular crystal is the only stable solid phase, the other crystals being metastable and sufficiently far above in enthalpy to be likely irrelevant for the kinetics of the fluid-to-solid transformation. We have then considered other interactions, smooth deformations of the PSM repulsion: the GEM potentials and the SVDW interaction. Looking at Table 1, where we collect the transition thresholds for all the cases considered, we see that a smoother interaction entails a higher transition pressure. Eventually, for $n=2$ (where $\widetilde{u}(k)$ is everywhere positive) crystallization is swept away at zero temperature.

We briefly comment about the possibility of a stable hexatic phase in a 2D quantum system at $T=0$, an issue that clearly goes beyond the scope of our mean-field analysis. To our knowledge, evidence of quasi-long-range bond-angle order in a quantum fluid has only been reported for distinguishable charges ($u(r)\propto 1/r$~\cite{Clark,Bruun}) and aligned dipoles ($u(r)\propto 1/r^3$~\cite{Lechner,Bruun}) confined in a plane. Both systems feature a hexatic phase in the classical regime ({\it i.e.}, for high temperature and/or large potential-to-kinetic energy ratio). When moderate quantum fluctuations are included, the hexatic phase is shifted to lower temperatures, while, deeper in the quantum regime, the hexatic phase is suppressed completely. It is not clear whether the hexatic order can survive down to zero temperature (evidences of opposite sign are given by Bruun and Nelson and by Clark {\it et al.}). We also underline that nothing precludes that hexatic order and cluster-crystal order can coexist in the same system, see for example \cite{Prestipino5}.

%
%
\begin{figure}
\begin{center}
\includegraphics[width=16cm]{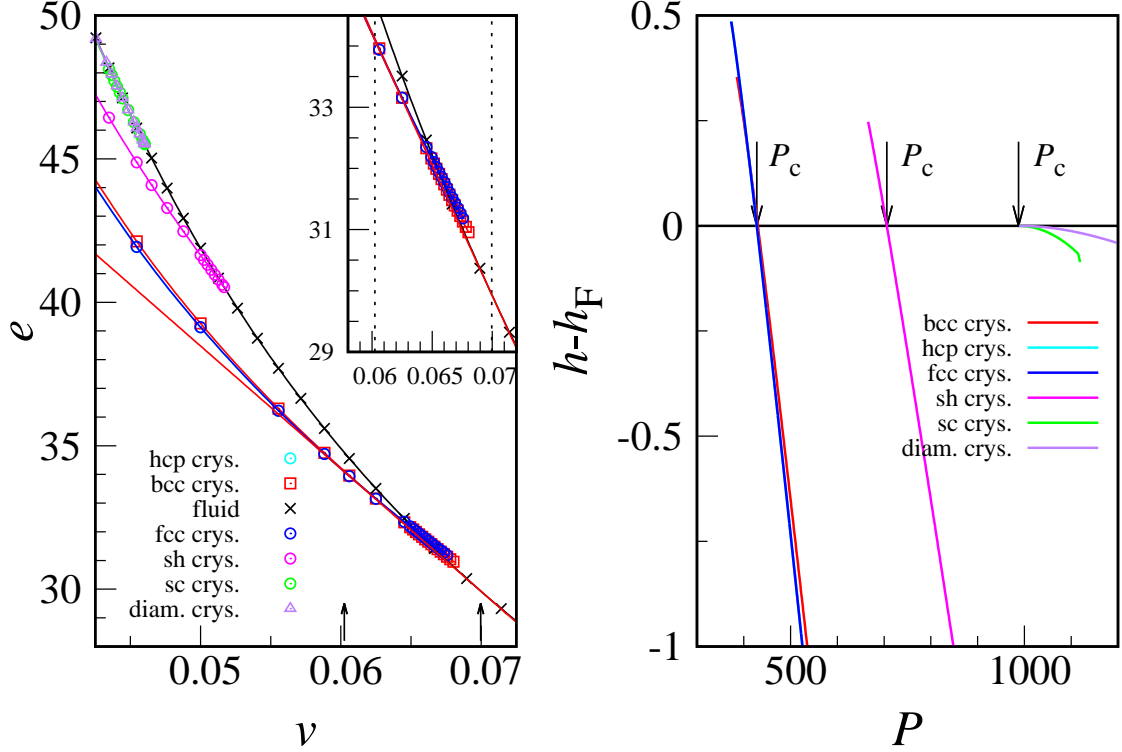}
\caption{PSM bosons in three dimensions at $T=0$, variational-theory results. Left: $e$ vs. $v$ for all the phases examined (see legend). The red straight line represents the common tangent to the fluid and fcc-crystal branches, whereas the arrows mark the location of the coexisting volumes. The inset shows a magnification of the transition region (the coexisting volumes are now signaled by two dotted lines). Right: enthalpy $h(P)$ for each solid phase, using the fluid phase as reference (black). The arrows mark the location of the transition into the fcc crystal (left), the sh crystal (middle), and the sc crystal (right). The transition pressure for the diamond crystal is the same as for the sc crystal.}
\label{fig6}
\end{center}
\end{figure}

%
%
\begin{figure}
\begin{center}
\includegraphics[width=12cm]{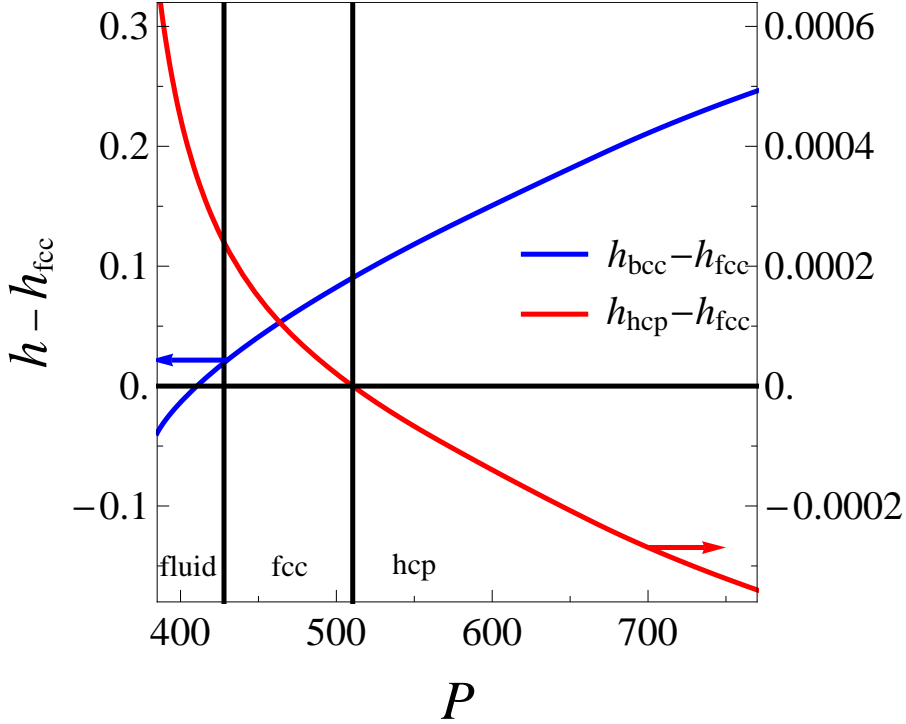}
\caption{PSM bosons in three dimensions at $T=0$: enthalpy $h(P)$ for the bcc (blue, left scale) and the hcp crystal (red, right scale), using the fcc crystal as reference. The vertical lines mark the location of the transitions (see Table 1).}
\label{fig7}
\end{center}
\end{figure}

%
%
\begin{figure}
\begin{center}
\includegraphics[width=16cm]{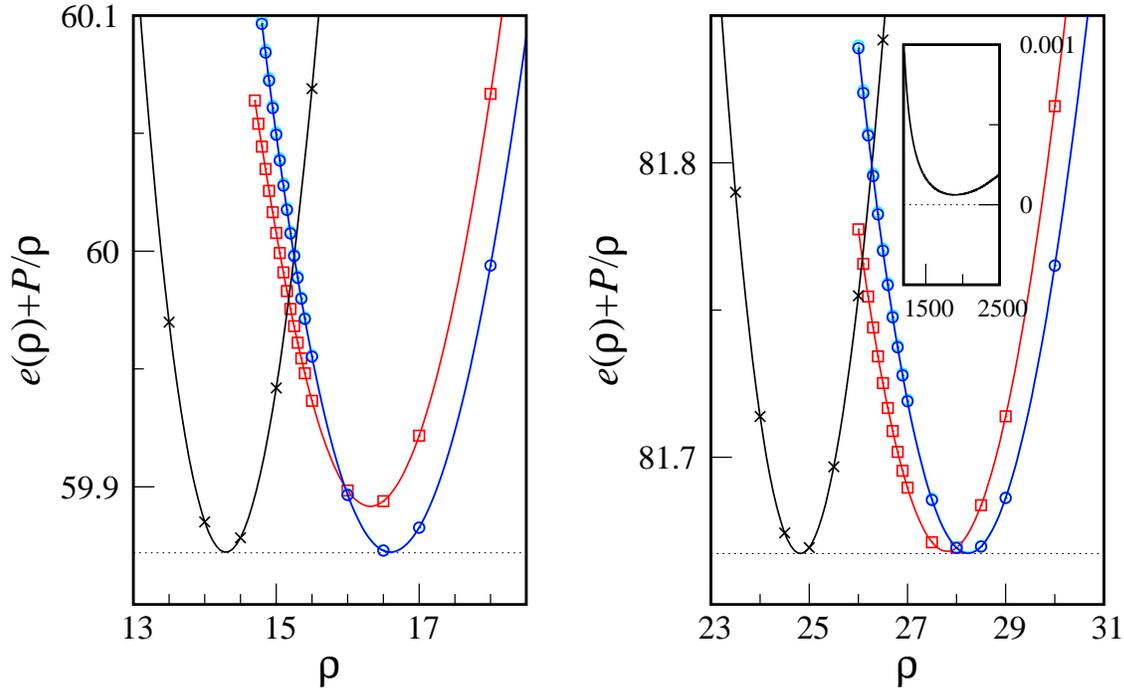}
\caption{Soft-core bosons in three dimensions at $T=0$: variational-theory results for the generalized enthalpy $\widetilde{h}(\rho)=e(\rho)+P/\rho$. Left: PSM, $P=427.89$ and $\mu=59.872$; right: SVDW, $P=1013.65$ and $\mu=81.667$. Besides fluid data (black crosses), results are reported for the fcc crystal (blue dots), the hcp crystal (cyan dots), and the bcc crystal (red dots). For both models freezing first occurs into a fcc crystal. In the inset of the right panel, we plot the difference in enthalpy between the hcp and the fcc crystal as a function of pressure. This difference is positive at all pressures, hence for $T=0$ the fcc crystal is always more stable than the hcp crystal.}
\label{fig8}
\end{center}
\end{figure}

%
%
\begin{figure}
\begin{center}
\includegraphics[width=16cm]{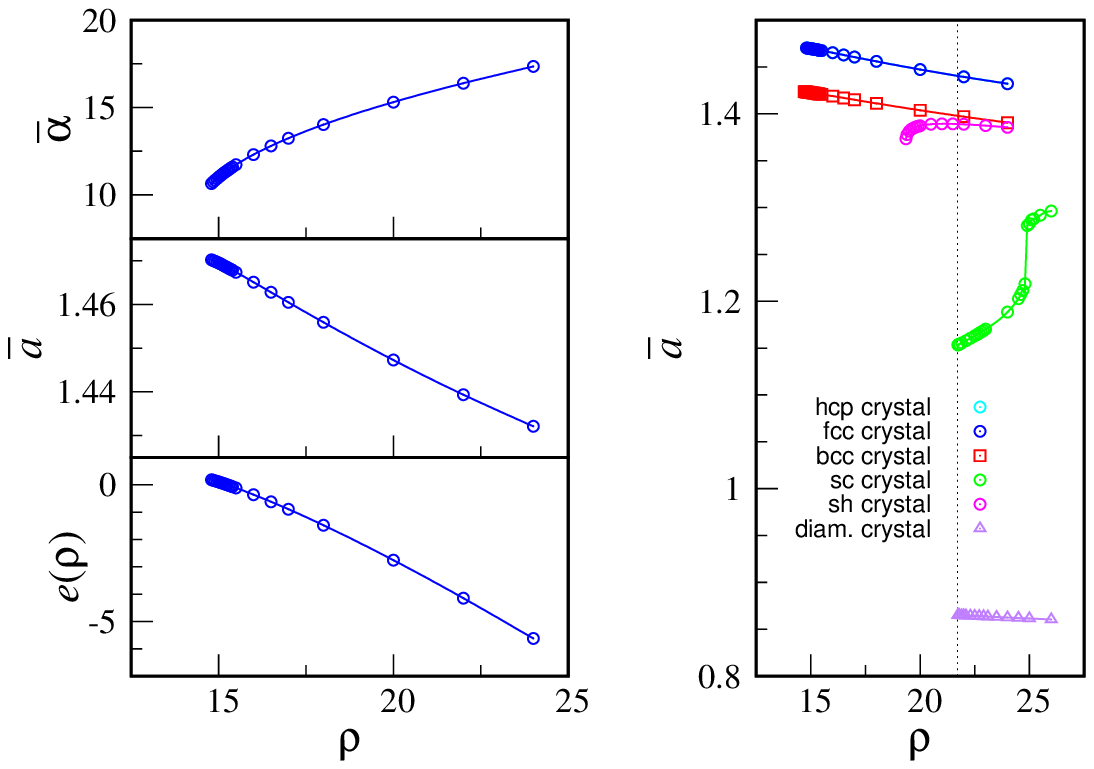}
\caption{PSM bosons in three dimensions at $T=0$. Left: optimal parameters for the fcc crystal as a function of the density according to Gaussian variational theory. Right: the lattice constant $\overline{a}$ is plotted as a function of $\rho$ for various 3D crystals (the hcp data are hidden behind fcc ones). A curious prediction of variational theory is the isostructural transition undergone by the sc crystal near $\rho=24.8$, signaled by a jump in the lattice constant and a cusp in the enthalpy (see Fig.\,6 right panel). The vertical dotted line marks continuous freezing ($\rho=21.71372\ldots$, see Appendix A).}
\label{fig9}
\end{center}
\end{figure}

Summarizing up to this point, MF results are confirmed by Gaussian variational theory both qualitatively and quantitatively. In particular, freezing in 2D occurs continuously for loosely-packed crystals, {\it i.e.}, those having a low coordination number $z$: the lower $z$ is, the smaller $\overline{a}$ in order to keep particles bound to each other. Furthermore, all crystals turn out to be cluster crystals: at the melting transition, the average number of particles per lattice site is spectacularly large and grows almost linearly with density (at melting, $\rho v_0$ is $25.98$ for the triangular crystal, $25.47$ for the square crystal, and $29.41$ for the honeycomb crystal).

In three dimensions, the competition for thermodynamic stability at $T=0$ is restricted to the fluid phase and the compact cubic phases only (fcc, bcc, and hcp), see Fig.\,6. Loosely-packed crystals, such as the simple-cubic (sc) crystal and the diamond crystal, melt continuously at a common critical pressure (see Appendix A.4 and A.5), much higher than the melting pressure of, say, the fcc phase. For PSM bosons, also the simple-hexagonal (sh) crystal is too far away in enthalpy from the fcc crystal to be of any relevance for crystallization (the $c/a$ ratio of the optimal sh crystal is approximately 0.90 near melting). The stable solid phase is the fcc crystal ($\rho_{\rm F0}=14.294,\rho_{\rm S0}=16.599$, and $\rho v_0=36.7$ at melting, in full agreement with the estimates in Ref.\,\cite{Ancilotto}), even though its enthalpy is only imperceptibly smaller than the hcp one (Fig.\,7). Upon increasing pressure, the hcp crystal eventually takes over, implying a solid-solid transition (transition thresholds can be read in Table 1). The bcc crystal, whose energy is lower than fcc energy at low density, is only metastable (see Fig.\,8 left panel). In Fig.\,9 the optimal values of the variational parameters are plotted as a function of $\rho$ for all phases. The situation is slightly different for SVDW bosons (Fig.\,8 right panel): the stable solid phase is now fcc at all pressures ($\rho_{\rm F0}=24.824,\rho_{\rm S0}=28.236$, and $\rho v_0=52.2$ at melting), but the bcc crystal (which is nearer in density to the fluid) is so close in enthalpy to the fcc crystal that, according to Ostwald's rule of stages~\cite{Ostwald}, the onset of the solid from the overcompressed fluid will occur through an initial stage characterized by a nucleus of prevailing bcc character~\cite{tenWolde,Auer,Chung,Levin}. This expectation is based on an analogy between quantum and classical (or thermal) nucleation, which we try to substantiate theoretically in Section IV. If Ostwald's rule applies~\cite{Sanders,Hedges}, the stable fcc structure will first appear in the core of near-critical nuclei, while bcc-like order survives in the external corona.

It is worth comparing the ground state of the quantum PSM and SVDW systems as a function of pressure to the phase diagram of the respective classical fluids. As far as the PSM is concerned, the fcc crystal is the only stable classical solid at low pressure~\cite{Zhang2}, exactly as in the case of PSM bosons at $T=0$. Instead, no phase diagram is available for classical SVDW particles. However, Zhang and Charbonneau~\cite{Zhang1} have reconstructed the 3D phase diagram of a similar system of classical particles interacting through the GEM-4 potential. In that case, cluster-crystal order at $T=0$ is fcc; but, at higher temperatures, the liquid first freezes into the cluster-bcc phase, which under pressure is eventually transformed into the cluster-fcc phase. This means that the $T=0$ chemical potential of the cluster-bcc crystal is close to that of the cluster-fcc crystal, {\it i.e.}, the same as found for weakly-repulsive SVDW bosons.

In Appendix B, we derive the MF spectrum of excitations in the fluid. This is accomplished by solving, in the amplitude-phase representation, the time-dependent GP equation for a slightly perturbed condensate wave function. The oscillatory solution obtained has the expected Bogoliubov-like dispersion~\cite{Macri2,Kunimi},
\be
\hbar\omega(k)=\sqrt{\frac{\hbar^2k^2}{2m}\left(\frac{\hbar^2k^2}{2m}+2\rho\widetilde{u}(k)\right)}\,.
\label{eq3-2}
\ee
If $\widetilde{u}(k)$ is negative in a range of $k$ values, the fluid phase exhibits superfluid behavior. Under the same assumption, $\omega(k)$ shows a roton minimum for sufficiently high density. The roton minimum becomes unstable exactly at the transition density for {\em continuous} freezing (see Appendix B), {\it i.e.}, at the upper threshold for thermodynamic stability of the fluid. Finally, we discuss in Appendix C the supersolid behavior of the crystal, {\it i.e.}, the property of anomalous rotational inertia~\cite{Leggett}. By adapting an argument exposed in Ref.\,\cite{Aftalion}, we find that within Gaussian variational theory any crystalline phase is necessarily supersolid at all pressures.

\section{The solid-fluid interface}
\setcounter{equation}{0}
\renewcommand{\theequation}{4.\arabic{equation}}

In this Section, we develop an elementary theory of the interface between solid and fluid at coexistence ($T=0$ and $P=P_{\rm coex}$). At a coarse-grained level of description the boundary region between the two phases is most easily represented in terms of the spatial dependence of a suitable order parameter distinguishing the two phases.

Let ${\cal E}_{\rm min}(\rho)$ represent the absolute minimum of the variational energy ${\cal E}(\alpha,a;\rho)$ as a function of $\rho$ (${\cal E}_{\rm min}(\rho)$ has a double-parabola shape, with a cusp at the point where the minimum jumps from $\alpha=0$ to $\alpha>0$). At a fixed pressure $P$, the equilibrium density $\rho_{\rm eq}(P)$ is the point of absolute minimum for ${\cal E}_{\rm min}(\rho)+P/\rho$, with $\rho_{\rm eq}(P)\equiv\rho_{\rm F}(P)<\rho_{\rm F0}$ for $P<P_{\rm coex}$ and $\rho_{\rm eq}(P)\equiv\rho_{\rm S}(P)>\rho_{\rm S0}$ for $P>P_{\rm coex}$; going across $P_{\rm coex}$, the equilibrium density jumps from $\rho=\rho_{\rm F0}$ to $\rho=\rho_{\rm S0}$. However, the fluid density is still defined above the coexistence pressure, at least up to $P^*$ (supercompressed fluid). Similarly, the solid density is also defined below the coexistence pressure, down to $P^{**}$ (undercompressed solid). In the interval $P^{**}<P<P^*$, which encloses $P_{\rm coex}$, $\rho_{\rm F}(P)$ and $\rho_{\rm S}(P)$ are both well defined.

At coexistence, solid and fluid have the same enthalpy:
{\small
\be
h_{\rm S}(P_{\rm coex})=h_{\rm F}(P_{\rm coex})\equiv\frac{\rho_{\rm F0}\widetilde{u}(0)}{2}+\frac{P_{\rm coex}}{\rho_{\rm F0}}\,.
\label{eq4-1}
\ee
}
In the interval $P^{**}<P<P^*$, the quantity (defined for every $\rho$)
\be
\Delta h(\rho;P,T=0)={\cal E}_{\rm min}(\rho)+\frac{P}{\rho}-h_{\rm F}(P)
\label{eq4-2}
\ee
has the shape of a double well, with two minima at $\rho_{\rm F}(P)$ and $\rho_{\rm S}(P)$, respectively equal to 0 and $h_{\rm S}(P)-h_{\rm F}(P)$. For $P=P_{\rm coex}$, the minima of $\Delta h$ are both zero, in agreement with Eq.\,(\ref{eq4-1}). For higher pressures, the high-density minimum (``solid'') is deeper than the low-density minimum (``fluid''), while the opposite occurs for $P<P_{\rm coex}$.

To describe the solid-fluid interface at coexistence, we promote the density $\rho$ to a field, $\rho({\bf x})$ (in every ``small'' region of solid, the value of $\alpha$ and $a$ will conform to the values expected for the bulk solid with density equal to the local one). In two dimensions, the density field at equilibrium will minimize the Landau free-energy functional~\cite{Prestipino3}
\be
{\cal H}[\rho]=\int{\rm d}^2x\left\{\frac{1}{2}c(\nabla\rho)^2+\frac{1}{2}\kappa(\nabla^2\rho)^2+h(\rho(x,y))\right\}\,,
\label{eq4-3}
\ee
where $c,\kappa>0$ are rigidity moduli and $h(\rho)\equiv N\Delta h(\rho;P_{\rm coex},T=0)/V$ is the enthalpy difference per unit volume between solid and fluid, so that ${\cal H}$ is the enthalpy content attached with the interface. The minimum of ${\cal H}$ must comply with boundary conditions. For example, if we want to describe a straight interface perpendicular to $x$, separating the solid (on the left) from the fluid (on the right), we should have:
\be
\rho(-\infty)=\rho_{\rm S0}\,\,\,\,\,\,{\rm and}\,\,\,\,\,\,\rho(+\infty)=\rho_{\rm F0}
\label{eq4-4}
\ee
(for symmetry reasons, $\rho$ will uniquely depend on $x$). Among all profiles that satisfy the conditions (\ref{eq4-4}), the equilibrium profile $\rho_0(x)$ minimizes
\be
{\cal H}[\rho]=L_y\int_{-\infty}^{+\infty}{\rm d}x\left\{\frac{1}{2}c\left(\frac{{\rm d}\rho}{{\rm d}x}\right)^2+\frac{1}{2}\kappa\left(\frac{{\rm d}^2\rho}{{\rm d}x^2}\right)^2+h(\rho(x))\right\}\,,
\label{eq4-5}
\ee
$L_y$ being the macroscopic transverse size of the sample. The value of ${\cal H}$ at the point of minimum is, by definition, $\gamma L_y$ ($\gamma$ is the {\em interface tension}). One finds~\cite{Prestipino3}:
\be
\gamma=\int_{-\infty}^{+\infty}{\rm d}x\left\{c\rho_0^{\prime 2}(x)+2\kappa\rho_0^{\prime\prime 2}(x)\right\}\,.
\label{eq4-6}
\ee
For $\kappa=0$ and $h(\rho)=h_0(1-\rho/\rho_{\rm F0})^2(1-\rho/\rho_{\rm S0})^2$ ($\phi^4$ theory) the Euler-Lagrange solution of (\ref{eq4-5}) is analytic~\cite{Barrat}:
\be
\rho_0(x)=\frac{\rho_{\rm S0}+\rho_{\rm F0}}{2}-\frac{\rho_{\rm S0}-\rho_{\rm F0}}{2}\tanh\frac{x}{\ell}
\label{eq4-7}
\ee
with $\ell=\rho_{\rm F0}\rho_{\rm S0}\sqrt{2c/h_0}/(\rho_{\rm S0}-\rho_{\rm F0})$, leading in turn to $\gamma=c(\rho_{\rm S0}-\rho_{\rm F0})^2/(3\ell)$. The limit of this approach to the description of the interface is that nobody knows how to extract the phenomenological coefficients $c$ and $\kappa$ from the microscopic interaction potential, hence $\gamma$ should actually be computed by another route.

The same functional (\ref{eq4-3}) can also serve to formulate in simple terms the process of solid nucleation from the fluid (assuming that a single order parameter suffices to characterize the solid cluster, see \cite{Prestipino4} for a discussion). Nucleation has to do with the decay of the fluid phase above $P_{\rm coex}$. This occurs through the onset of a sufficiently large solid inclusion, or cluster, defining the enthalpy barrier that should be overcome in order for crystallization to occur.  However, at variance with thermally-activated nucleation, the overcoming of the nucleation barrier here occurs at $T=0$, triggered by quantum fluctuations alone (see, e.g., Ref.\,\cite{Tsymbalenko} and references cited therein).

We briefly review the derivation of the cluster free energy proposed in Ref.\,\cite{Prestipino3}. While at coexistence $h(\rho)$ has two minima of equal depth, for pressures higher than $P_{\rm coex}$ the absolute minimum of $h(\rho)$ falls at $\rho=\rho_{\rm S}>\rho_{\rm S0}$. At an elementary level, this function can be represented as a fourth-order polynomial,
\be
h(\rho)=c_2\rho^2+c_3\rho^3+c_4\rho^4
\label{eq4-8}
\ee
with $c_2=c_{20}-c_{20}'(P-P_{\rm coex})$ ($c_{20},c_{20}'>0$), all other $c_n$ being constant. Near $P_{\rm coex}$, the spatial profile of the order parameter for a spherical solid cluster of radius $R\gg\sqrt{2c/c_{20}}$ is well described by $\rho_0(r-R)$ (providing that the ``center'' of $\rho_0(x)$ is chosen at $x=0$). In this case, the free-energy cost for the cluster becomes:
{\small
\be
\Delta G(R)=2\pi\int_0^{+\infty}{\rm d}r\,r[c\rho_0^{\prime 2}(r-R)+2\kappa\rho_0^{\prime\prime 2}(r-R)]-2\pi c_{20}'\Delta P\int_0^{+\infty}{\rm d}r\,r\rho_0^2(r-R)\,.
\label{eq4-9}
\ee
}
Following the same steps as in Ref.\,\cite{Prestipino3}, we eventually arrive at the following MF expression of the cost of cluster formation:
\be
\Delta G(R)=2\pi R\widetilde{\gamma}\left(1-\frac{2\widetilde{\delta}}{R}+\frac{\widetilde{\epsilon}}{R^2}\right)-\pi R^2\rho_{\rm S0}|\Delta\mu|
\label{eq4-10}
\ee
with $\Delta\mu=-c_{20}'\rho_{\rm S0}\Delta P<0$ and $\widetilde{\gamma},\widetilde{\delta},\widetilde{\epsilon}$ linear functions of $\Delta P$. Equation (\ref{eq4-10}) is similar to the MF cost of cluster formation for thermal nucleation~\cite{Fisher,Prestipino3}. At coexistence, while $\widetilde{\gamma}$ reduces to $\gamma$ (Eq.\,(\ref{eq4-6})), $\widetilde{\delta}$ becomes
\be
\delta=-\frac{\int_{-\infty}^{+\infty}{\rm d}x\,x\left(c\rho_0^{\prime 2}+2\kappa\rho_0^{\prime\prime 2}\right)}{\int_{-\infty}^{+\infty}{\rm d}x\left(c\rho_0^{\prime 2}+2\kappa\rho_0^{\prime\prime 2}\right)}\,.
\label{eq4-11}
\ee
The value of $\delta$ (``Tolman's length'') is non-zero if $\rho_0(x)$ is an asymmetric density profile, like in case of an interface between phases of different nature.

\section{Conclusions}

In this paper we employ MF theory to study pressure-driven crystallization of soft-core bosons at $T=0$, in two and three dimensions. Within this theory, the ground state of the system is represented as a perfect condensate, which is realistic for weak interparticle forces (ultracold atomic gases can approach this condition closely). However, rather than solving the GP equation, which is tantamount to selecting the best MF state, we make a two-parameter ansatz on the single-particle wave function which has the advantage of speeding up calculations considerably, without affecting accuracy in any sensible way. This is especially true in three dimensions, where obtaining self-consistency in the GP equation is painfully slow.

By means of the variational method we compute the energy of many crystalline states, then deciding which phase is stable at the given pressure by a comparison of their enthalpies. As a rule, these crystals are cluster crystals, meaning that site occupancy is larger than one. Moreover, they are supersolids, meaning that the moment of inertia is diminished with respect to a classical solid. In two dimensions, the best crystalline ground state is triangular, and the freezing transition is first-order. On other lattices (square, honeycomb, and striped) freezing would be continuous and, more importantly, it will occur at the same pressure for all; this critical pressure also corresponds to the highest pressure at which the fluid can exist as a superfluid (these features also hold in three dimensions). We find that crystallization is pushed to higher and higher pressures when the exponent $n$ in the GEM potential is reduced from $n=\infty$ (PSM limit), until freezing is completely washed out in the Gaussian, $n=2$ case.

The phase diagram in three dimensions is more crucially dependent on the nature of the interaction. While open, low-coordinated crystals (like simple-cubic and diamond crystals) are always metastable, the enthalpies of the other cubic crystals are close to each other. For PSM bosons the phase sequence for increasing pressure is fluid-fcc-hcp; for a softened van der Waals repulsion the only stable crystal is fcc (even though close to freezing the bcc crystal is only marginally less stable than the fcc crystal).

From a more general perspective, our results confirm the idea of relating the supersolid state of soft-core bosons to the clustering behavior of the solid: in essence, within a MF approximation the quantum theory can be mapped onto a classical-like description in terms of quantum densities~\cite{Pomeau}, which makes the classical analog more apparent. It is also by virtue of such a quantum-to-classical mapping, realized through the variational approach, that we have shown that quantum nucleation of the solid from the fluid can be treated along the same lines of the better known process of nucleation induced by thermal fluctuations.

In a forthcoming publication, we will examine in depth the case of one-dimensional soft-core bosons (which can be realized in elongated optical or magnetic traps with Rydberg-dressed atoms), where the reconstruction of phase diagram by means of Gaussian variational theory is to a large extent fully analytic.

\appendix
\section{When freezing is continuous}
\renewcommand{\theequation}{A.\arabic{equation}}

As discussed in Section III, on some lattices the freezing transition of soft-core bosons at $T=0$ turns out to be continuous. In this event, a low-$\alpha$ expansion of the energy functional ${\cal E}(\alpha,a;\rho)$ allows one to compute the transition point exactly. We illustrate in some detail how this expansion is worked out for the square lattice, while we only state results for other lattices.

\subsection{Square lattice}

We first use Eq.\,(\ref{eq2-13}) to develop, on the basis of Eq.\,(\ref{eq2-17}), a low-$\alpha$ kinetic-energy expansion. By ordering terms according to their relative importance for small $\alpha$ values, we readily obtain:
\be
{\cal E}_{\rm kin}=\frac{8\pi^2\sigma^2}{a^2}e_0\left(X^2-2X^4+\ldots\right)\,\,\,\,\,\,{\rm with}\,\,\,\,\,\,X=e^{-\pi^2/\alpha}\,.
\label{a-1}
\ee
As for the potential energy, we should estimate all the sums appearing in Eqs.\,(\ref{eq2-21}). In this case too, $X$ proves to be a natural expansion variable. The derivation is straightforward but lengthy; the final result is:
{\small
\ba
{\cal E}_{\rm pot}&=&\frac{\rho}{2}\left\{4(\widetilde{u}(-1,0)+\widetilde{u}(0,-1)+\widetilde{u}(1,0)+\widetilde{u}(0,1))X^2\right.
\nonumber \\
&+&\left[\widetilde{u}(-2,0)+\widetilde{u}(0,-2)+\widetilde{u}(2,0)+\widetilde{u}(0,2)-16(\widetilde{u}(-1,0)+\widetilde{u}(0,-1)+\widetilde{u}(1,0)+\widetilde{u}(0,1))\right.
\nonumber \\
&+&\left.\left.16(\widetilde{u}(-1,-1)+\widetilde{u}(-1,1)+\widetilde{u}(1,-1)+\widetilde{u}(1,1))\right]X^4+\ldots\right\}\,,
\label{a-2}
\ea
}
where $\widetilde{u}(-1,0)$ stands for $\widetilde{u}(G)$ with ${\bf G}=(-1){\bf b}_1+0{\bf b}_2$, and so on. Putting together (\ref{a-1}) and (\ref{a-2}), we obtain the following expansion for the excess energy of the crystal:
\ba
&&\Delta{\cal E}\equiv{\cal E}_{\rm cin}+{\cal E}_{\rm pot}-\frac{\rho\widetilde{u}(0)}{2}=8\left(\frac{\pi^2\sigma^2}{a^2}e_0+\rho\widetilde{u}\left(\frac{2\pi}{a}\right)\right)X^2
\nonumber \\
&&+2\left(-\frac{8\pi^2\sigma^2}{a^2}e_0-16\rho\widetilde{u}\left(\frac{2\pi}{a}\right)+16\rho\widetilde{u}\left(\frac{2\sqrt{2}\pi}{a}\right)+\rho\widetilde{u}\left(\frac{4\pi}{a}\right)\right)X^4+\ldots
\label{a-3}
\ea
The extremal points of $\Delta{\cal E}\equiv rX^2+wX^4$ are the roots of $\Delta{\cal E}'(\alpha)=0$, that is $\overline{X}=0$ and (provided $r<0$ and $w>0$) $\overline{X}=\sqrt{-r/(2w)}$, with specific energies equal to $\Delta{\cal E}=0$ and $\Delta{\cal E}=-r^2/(4w)$, respectively. To be specific, let us consider the PSM case, for which $\widetilde{u}(0)=\pi\epsilon\sigma^2$ and $\widetilde{u}(k)=2\pi\epsilon\sigma J_1(k\sigma)/k$ for $k>0$ ($J_1$ is a Bessel function of the first kind). A non-trivial solution to $\Delta{\cal E}'(\alpha)=0$ exists when $r<0$, that is $\rho>\rho_0(a)$ with
\be
\rho_0\sigma^2=-\frac{e_0}{\epsilon}\frac{\pi^2}{\left(\frac{a}{\sigma}\right)^3J_1\left(\frac{2\pi}{a}\sigma\right)}\,.
\label{a-4}
\ee
The density $\rho_0$ is positive for $0.89560368\ldots<a/\sigma<1.63978795\ldots$ (we note that $J_1(2\pi\sigma/a)<0$ also in other ranges of $a$, but the corresponding energy extrema are non-optimal). The smallest $\rho_0$ at which $\Delta{\cal E}$ turns negative is the minimum of $\rho_0(a)$ in the above interval, that is the maximum of $y=-x^3J_1(2\pi/x)$. The derivative $y'$ is positive for
\be
x<1.31474663\ldots\equiv\frac{a_c}{\sigma}\,.
\label{a-5}
\ee
Hence, the transition occurs for $\rho=\rho_0(a_c)=14.73710\ldots\,e_0\epsilon^{-1}\sigma^{-2}\equiv\rho_c$ ($w$ is strictly positive near $\rho_c$); the $a$ and $\rho$ values at the transition are fully consistent with the numerical solution (see Fig.\,3 right panel). At $\rho_c$, $\overline{X}$ switches continuously from 0 (fluid) to a value $\propto(\rho-\rho_c)/\rho_c$ (crystal). Right at the transition, $P=P_c=\pi\epsilon\rho_c^2\sigma^2/2=341.1488\ldots\,e_0^2\epsilon^{-1}\sigma^{-2}$. Slightly above $\rho_c$, where $r\simeq r_0(1-\rho/\rho_c)$ and $w\simeq w_0>0$, the optimal $\alpha$ value and the excess energy behave as:
\be
\overline{\alpha}(\rho)\sim\frac{2\pi^2}{\left|\ln\left(\frac{r_0}{2w_0}\frac{\rho-\rho_c}{\rho_c}\right)\right|}\,\,\,\,\,\,{\rm and}\,\,\,\,\,\,\Delta e(\rho)\sim -\frac{r_0^2}{4w_0\rho_c^2}(\rho-\rho_c)^2\,;
\label{a-6}
\ee
moreover, the average number of particles per cluster equals $\rho_ca_c^2\simeq 25.4739\ldots\,e_0/\epsilon$, which is surprisingly large. Finally, from the general formula of the isothermal compressibility,
\be
K_T^{-1}=-\left.V\frac{\partial P}{\partial V}\right|_{T=0}=\rho P'(\rho)=2\rho^2 e'(\rho)+\rho^3 e''(\rho)\,,
\label{a-7}
\ee
it follows that $K_T^{-1}$ has different values in the two phases for $\rho=\rho_c$:
\be
{\rm F}:\,\,\,K_T^{-1}=\pi\epsilon\rho_c^2\sigma^2\,;\,\,\,\,\,\,\,\,\,\,{\rm S}:\,\,\,K_T^{-1}=\pi\epsilon\rho_c^2\sigma^2-\frac{r_0^2}{2w_0}\rho_c^3\,.
\label{a-8}
\ee
Therefore, $K_T$ shows a jump at the transition and, exactly at $\rho_c$, the solid is {\em more} compressible than the fluid. 

\subsection{Stripes}

A periodic one-dimensional modulation of the single-particle wave function in 2D corresponds to a striped crystal. Denoting $L_x$ the macroscopic size of the lattice in the direction of system periodicity, and $L_y$ the size of the sample in the perpendicular direction, the variational wave function reads:
\be
\psi({\bf x})=\frac{1}{\sqrt{L_x}}\sum_G\psi_Ge^{iGx}\cdot\frac{1}{\sqrt{L_y}}\equiv\psi_{0x}(x)\psi_{0y}
\label{a-9}
\ee
with $G=(2\pi/a)n$ ($a$ is the periodicity along $x$ and $n$ is any integer). Moreover, 
\be
\psi_G=\left(\frac{2\pi}{\alpha I(\alpha)^2}\right)^{1/4}e^{-\frac{G^2a^2}{4\alpha}}\,\,\,\,\,\,{\rm with}\,\,\,\,\,\,I(\alpha)=\sum_{n=-\infty}^{+\infty}e^{-\frac{\alpha}{2}n^2}\,.
\label{a-10}
\ee
Plugging (\ref{a-9}) into the Hartree energy functional, we obtain:
\be
{\cal E}=\frac{1}{2}e_0\sum_G(G\sigma)^2\psi_G^2+\frac{1}{2}\rho\sum_{G_1,G_2,G_3}\widetilde{u}(|G_1|)\psi_{G_1+G_2}\psi_{G_1+G_3}\psi_{G_2}\psi_{G_3}\,,
\label{a-11}
\ee
where $\widetilde{u}(|G_1|)$ is the Fourier transform of the 2D potential computed in $|G_1|$. A derivation similar to that worked out for the square lattice then leads to $\Delta{\cal E}\equiv rX^2+wX^4+\ldots$ with
\be
r=4\left(\frac{\pi^2\sigma^2}{a^2}e_0+\rho\widetilde{u}\left(\frac{2\pi}{a}\right)\right)\,\,\,\,\,\,{\rm and}\,\,\,\,\,\,w=-\frac{8\pi^2\sigma^2}{a^2}e_0+\rho\left(\widetilde{u}\left(\frac{4\pi}{a}\right)-16\widetilde{u}\left(\frac{2\pi}{a}\right)\right)\,.
\label{a-12}
\ee
Note, in particular, that the $X^2$ term changes sign at the same density $\rho_0(a)$ as for the square lattice, hence the considerations made for the square crystal also apply for stripes. In particular, for the PSM we find (using reduced units) $a_c=1.31474\ldots,\rho_c=14.73710\ldots$, and $P_c=341.1488\ldots$

\subsection{Honeycomb lattice}

For the honeycomb lattice, the excess energy reads:
\ba
\Delta{\cal E}&\equiv&{\cal E}_{\rm cin}+{\cal E}_{\rm pot}-\frac{\rho\widetilde{u}(0)}{2}
\nonumber \\
&=&\left(\frac{4\pi^2\sigma^2}{3a^2}e_0+3\rho\widetilde{u}\left(\frac{4\pi}{3a}\right)\right)e^{-\frac{8\pi^2}{9\alpha}}-3\rho\widetilde{u}\left(\frac{4\pi}{3a}\right)e^{-\frac{4\pi^2}{3\alpha}}+\ldots
\label{a-13}
\ea
In the PSM case, the coefficient of the leading term in the low-$\alpha$ expansion of $\Delta{\cal E}$ changes sign at the density
\be
\rho_0\sigma^2=-\frac{e_0}{\epsilon}\frac{8\pi^2}{27\left(\frac{a}{\sigma}\right)^3J_1\left(\frac{4\pi}{3a}\sigma\right)}\,.
\label{a-14}
\ee
In order that $\rho_0(a)>0$, it must be $0.59706912\ldots<a<1.09319196\ldots$ (in the same interval the subleading term in Eq.\,(\ref{a-13}) is positive). The smallest $\rho$ above which the energy becomes negative is the minimum of $\rho_0(a)$. For $y=-x^3J_1(4\pi/(3x))$, the derivative $y'$ is positive for
\be
x<0.876498\ldots\equiv a_c=\frac{2}{3}\cdot 1.31474663\ldots
\label{a-15}
\ee
Hence, the transition is continuous and occurs exactly at the same density $\rho_c=14.73710\ldots$ of the square lattice; also the critical pressure is the same: $P_c=\pi\epsilon\rho_c^2\sigma^2/2=341.1488\ldots$

\subsection{Simple-cubic lattice}

Numerical analysis suggests that crystallization is continuous also on the simple-cubic lattice. Indeed, for small $\alpha$ we have:
{\small
\ba
&&\Delta{\cal E}\equiv{\cal E}_{\rm cin}+{\cal E}_{\rm pot}-\frac{\rho\widetilde{u}(0)}{2}=12\left(\frac{\pi^2\sigma^2}{a^2}e_0+\rho\widetilde{u}\left(\frac{2\pi}{a}\right)\right)X^2
\nonumber \\
&&+3\left(-\frac{8\pi^2\sigma^2}{a^2}e_0-16\rho\widetilde{u}\left(\frac{2\pi}{a}\right)+32\rho\widetilde{u}\left(\frac{2\sqrt{2}\pi}{a}\right)+\rho\widetilde{u}\left(\frac{4\pi}{a}\right)\right)X^4+\ldots
\label{a-16}
\ea
}
For the PSM, where $\widetilde{u}(k)=4\pi\epsilon[\sin(k\sigma)-k\sigma\cos(k\sigma)]/k^3$ and $\widetilde{u}(0)=4\pi\epsilon\sigma^3/3$, the coefficient of $X^2$ changes sign at
\be
\rho_0\sigma^3=-\frac{e_0}{\epsilon}\frac{2\pi^4}{\left(\frac{a}{\sigma}\right)^5\left(\sin\frac{2\pi\sigma}{a}-\frac{2\pi\sigma}{a}\cos\frac{2\pi\sigma}{a}\right)}\,.
\label{a-17}
\ee
The smallest $\rho$ value above which the energy becomes negative is the minimum of $\rho_0(a)$. For $y=-x^5[\sin(2\pi/x)-(2\pi/x)\cos(2\pi/x)]$, the derivative $y'>0$ for
\be
x<1.15317\ldots\equiv a_c
\label{a-18}
\ee
Hence, the transition occurs at $\rho_c=\rho_0(a_c)=21.71372\ldots$ (reduced units). Near this density, the coefficient of $X^4$ is positive. The transition pressure is $P_c=2\pi\epsilon\sigma^3\rho_c^2/3=987.4772\ldots$

\subsection{Diamond lattice}

The diamond lattice can be described as a fcc lattice with a two-atom basis:
\ba
&&{\bf a}_1=(2a/\sqrt{3})(0,1,1)\,,\,\,\,{\bf a}_2=(2a/\sqrt{3})(1,0,1)\,,\,\,\,{\bf a}_3=(2a/\sqrt{3})(1,1,0)\,;
\nonumber \\
&&{\bf e}_1=(0,0,0)\,\,\,\,\,\,{\rm and}\,\,\,\,\,\,{\bf e}_2=(a/\sqrt{3})(1,1,1)\,,
\label{a-19}
\ea
again denoting $a$ the nearest-neighbor distance. The volume of the primitive cell is $v_0=|{\bf a}_1\cdot{\bf a}_2\wedge{\bf a}_3|=16/(3\sqrt{3})a^3$, whereas the reciprocal-lattice vectors are:
\be
{\bf b}_1=\frac{\sqrt{3}\pi}{2a}(-1,1,1)\,,\,\,\,{\bf b}_2=\frac{\sqrt{3}\pi}{2a}(1,-1,1)\,,\,\,\,\,\,\,{\rm and}\,\,\,\,\,\,{\bf b}_3=\frac{\sqrt{3}\pi}{2a}(1,1,-1)\,.
\label{a-20}
\ee
On this lattice too the freezing transition of the PSM is continuous and falls at the density $\rho_c=21.71372\ldots$ Indeed,
\ba
&&\Delta{\cal E}\equiv{\cal E}_{\rm cin}+{\cal E}_{\rm pot}-\frac{\rho\widetilde{u}(0)}{2}=8\left(\frac{\pi^2\sigma^2}{a^{\prime 2}}e_0+\rho\widetilde{u}\left(\frac{2\pi}{a'}\right)\right)e^{-\frac{9\pi^2}{8\alpha}}
\nonumber \\
&&+\left(-\frac{18\pi^2\sigma^2}{a^2}e_0+\rho\widetilde{u}\left(\frac{3\pi}{a}\right)+6\rho\widetilde{u}\left(\frac{\sqrt{6}\pi}{a}\right)-64\rho\widetilde{u}\left(\frac{3\pi}{2a}\right)\right)e^{-\frac{9\pi^2}{4\alpha}}+\ldots
\label{a-21}
\ea
with $a'=(4/3)a$. The lowest density $\rho_c$ at which the coefficient of the leading term in (\ref{a-21}) becomes negative is the same as for the simple-cubic lattice. The transition pressure is also the same, $P_c=987.4772\ldots$ However, the value of $a$ at the transition is $3/4$ of that in (\ref{a-18}), namely $0.86487\ldots$, as fully confirmed by numerical calculations.

\section{Spectrum of excitations in the fluid}
\renewcommand{\theequation}{B.\arabic{equation}}

In order to investigate the collective excitations of the system in the fluid phase, one possibility is to solve the so-called Bogoliubov-De Gennes equations, as illustrated e.g. in Ref.\,\cite{Macri1}. We here follow a different route, by elaborating on an argument in \cite{Nore}. The starting point is time-dependent GP equation~\cite{Gross1,Pitaevskii,Gross2},
\be
i\hbar\frac{\partial\psi}{\partial t}({\bf x},t)=-\frac{\hbar^2}{2m}\nabla^2\psi({\bf x},t)+N\int{\rm d}^dy\,|\psi({\bf y},t)|^2u({\bf x}-{\bf y})\psi({\bf x},t)\,,
\label{b-1}
\ee
which describes the MF dynamics of a system of identical bosons at $T=0$. This equation is the bosonic variant of the time-dependent Hartree equations, in turn derived from the quantum variational principle once the action has been specialized to a factorized $t$-dependent system ground state, in the same spirit of the ansatz (\ref{eq2-2}). Multiplying Eq.\,(\ref{b-1}) by $\psi^*$ and then subtracting the complex conjugate of the resulting equation, we arrive at
\be
\frac{\partial}{\partial t}(\psi^*\psi)+\frac{i\hbar}{2m}\nabla\cdot(\psi\nabla\psi^*-\psi^*\nabla\psi)=0\,,
\label{b-2}
\ee
which has the form of a continuity equation if a velocity field is defined by
\be
{\bf v}=\frac{i\hbar}{2m}\frac{\psi\nabla\psi^*-\psi^*\nabla\psi}{|\psi|^2}\,.
\label{b-3}
\ee
Indeed, writing $\psi$ in terms of its amplitude and phase,
\be
\psi=\frac{1}{\sqrt{V}}\sqrt{\eta({\bf x},t)}e^{i\theta({\bf x},t)}\,,
\label{b-4}
\ee
Eq.\,(\ref{b-3}) is rewritten as
\be
\frac{\partial\eta}{\partial t}+\nabla\cdot(\eta{\bf v})=0\,\,\,\,\,\,{\rm with}\,\,\,\,\,\,{\bf v}=\frac{\hbar}{m}\nabla\theta\,.
\label{b-5}
\ee
However, the previous equation is not sufficient to calculate both $\eta$ and $\theta$. Another equation can be obtained by plugging Eq.\,(\ref{b-4}) directly into the GP equation, with the result that:
\ba
&&i\hbar\frac{1}{2\sqrt{\eta}}\frac{\partial\eta}{\partial t}-\hbar\sqrt{\eta}\frac{\partial\theta}{\partial t}=-\frac{\hbar^2}{2m}\frac{i}{\sqrt{\eta}}\nabla\eta\cdot\nabla\theta-\frac{\hbar^2}{2m}i\sqrt{\eta}\nabla^2\theta
\nonumber \\
&+&\frac{\hbar^2}{8m}\frac{(\nabla\eta)^2}{\eta^{3/2}}-\frac{\hbar^2}{4m}\frac{\nabla^2\eta}{\sqrt{\eta}}+\frac{\hbar^2}{2m}\sqrt{\eta}(\nabla\theta)^2+\rho\int{\rm d}^dy\,\eta({\bf y},t)u({\bf x}-{\bf y})\sqrt{\eta}\,.
\label{b-6}
\ea
While the imaginary part of (\ref{b-6}) reproduces the continuity equation (\ref{b-5}), the real part reads:
\be
-\hbar\frac{\partial\theta}{\partial t}=-\frac{\hbar^2}{2m}\frac{\nabla^2\sqrt{\eta}}{\sqrt{\eta}}+\frac{\hbar^2}{2m}(\nabla\theta)^2+\rho\int{\rm d}^dy\,\eta({\bf y},t)u({\bf x}-{\bf y})\,.
\label{b-7}
\ee
Taking the gradient of (\ref{b-7}), the outcome is:
\be
m\frac{\partial{\bf v}}{\partial t}+m({\bf v}\cdot\nabla){\bf v}=\frac{\hbar^2}{2m}\nabla\left(\frac{\nabla^2\sqrt{\eta}}{\sqrt{\eta}}\right)-\rho\nabla\int{\rm d}^dy\,\eta({\bf y},t)u({\bf x}-{\bf y})\,.
\label{b-8}
\ee
Equation (\ref{b-8}) resembles a Navier-Stokes equation without viscosity term, if we identify the first term on the right-hand side with (minus) a pressure gradient.

The final step consists in deriving an equation for the small deviations of $\eta$ from the homogeneous-fluid solution $\eta=1$ and $\nabla\theta=0$. Such perturbed solutions are the sought-for excited states. Inserting $\eta=1+\delta\eta$ and $\nabla\theta=\delta{\bf u}$ into Eqs.\,(\ref{b-5}) and (\ref{b-8}), and simply ignoring every term that is not linear in $\delta\eta$ or $\delta{\bf u}$, we get from the continuity equation:
\be
\frac{\partial\delta\eta}{\partial t}+\frac{\hbar}{m}\nabla\cdot\delta{\bf u}=0\,\,\,\,\,\,\Longrightarrow\,\,\,\,\,\,\frac{\partial^2\delta\eta}{\partial t^2}=-\frac{\hbar}{m}\nabla\cdot\left(\frac{\partial\delta{\bf u}}{\partial t}\right)\,.
\label{b-9}
\ee
Moreover, we have:
\ba
m\frac{\partial{\bf v}}{\partial t}&=&\hbar\frac{\partial\delta{\bf u}}{\partial t}\,;
\nonumber \\
m({\bf v}\cdot\nabla){\bf v}&=&\frac{m}{2}\nabla(v^2)=\frac{\hbar^2}{2m}\nabla(\delta u^2)={\cal O}(\delta u^2)\,;
\nonumber \\
\frac{\hbar^2}{2m}\nabla\left(\frac{\nabla^2\sqrt{\eta}}{\sqrt{\eta}}\right)&=&\frac{\hbar^2}{4m}\nabla(\nabla^2\delta\eta)\,;
\nonumber \\
-\rho\nabla\int{\rm d}^dy\,\eta({\bf y},t)u({\bf x}-{\bf y})&=&-\rho\nabla\int{\rm d}^dy\,\delta\eta({\bf y},t)u({\bf x}-{\bf y})\,,
\label{b-10}
\ea
which eventually simplify Eq.\,(\ref{b-8}) to:
\be
\frac{\partial\delta{\bf u}}{\partial t}=\frac{\hbar}{4m}\nabla(\nabla^2\delta\eta)-\frac{1}{\hbar}\rho\nabla\int{\rm d}^dy\,\delta\eta({\bf y},t)u({\bf x}-{\bf y})\,.
\label{b-11}
\ee
Inserting Eq.\,(\ref{b-11}) into the second of Eqs.\,(\ref{b-9}), we finally obtain:
\be
\frac{\partial^2\delta\eta}{\partial t^2}=-\frac{\hbar^2}{4m^2}\nabla^2(\nabla^2\delta\eta)+\frac{1}{m}\rho\nabla^2\int{\rm d}^dy\,\delta\eta({\bf y},t)u({\bf x}-{\bf y})\,.
\label{b-12}
\ee
This equation admits solutions in the form of plane waves, $\delta\eta=\varepsilon\cos({\bf k}\cdot{\bf x}-\omega t)$, where $\varepsilon$ is a small dimensionless amplitude. The dispersion relation of these waves can be obtained by observing that:
\ba
&&\frac{\partial^2\delta\eta}{\partial t^2}=-\omega^2\delta\eta\,,\,\,\,\nabla^2\delta\eta=-k^2\delta\eta\,,\,\,\,\nabla^2(\nabla^2\delta\eta)=k^4\delta\eta\,,\,\,\,{\rm and}
\nonumber \\
&&\nabla^2\int{\rm d}^dy\,\delta\eta({\bf y},t)u({\bf x}-{\bf y})=-k^2\widetilde{u}(k)\delta\eta\,.
\label{b-13}
\ea
Substituting these formulae into Eq.\,(\ref{b-12}), we finally arrive at:
\be
\hbar^2\omega^2=\frac{\hbar^2k^2}{2m}\left(\frac{\hbar^2k^2}{2m}+2\rho\widetilde{u}(k)\right)\,,
\label{b-14}
\ee
which coincides with Eq.\,(6) of Ref.\,\cite{Macri1} and is identical (for a contact interaction) to the celebrated Bogoliubov spectrum. As long as the r.h.s. of (\ref{b-14}) is positive, and $\widetilde{u}(k)$ is negative in some range of $k$, the fluid is (by Landau's argument) {\em superfluid}. We have recently become aware of a different approach to obtain the excitation spectrum of a superfluid, based on the use of the non-linear logarithmic Schr\"odinger equation\,\cite{Bialynicki-Birula,Avdeenkov,Zloshchastiev}.

For the PSM in 2D, a roton minimum is only present in the excitation spectrum if the density is larger than $5.032$; moreover, the r.h.s. of (\ref{b-14}) is positive up to $g\equiv\rho\widetilde{u}(0)/e_0=46.2979\ldots$, corresponding to exactly the same density $14.7371\ldots$ where continuous freezing occurs (crystallization into a triangular crystal occurs at a density {\em smaller} than this). The behavior is analogous in 3D: The r.h.s. of Eq.\,(\ref{b-14}) is positive up to $\rho\widetilde{u}(0)/e_0=90.9542\ldots$, which corresponds to the same density ($21.7137\ldots$) where continuous freezing takes place. In fact, this result is absolutely general. Using $k=2\pi/a$ in the quantity within parentheses in (\ref{b-14}), we see that it reduces to
\be
2\left(\frac{\pi^2\sigma^2}{a^2}e_0+\rho\widetilde{u}\left(\frac{2\pi}{a}\right)\right)\,,
\label{b-15}
\ee
which, up to a constant factor, is identical to the quantity $r$ encountered in Appendix A, whose crossover from positive to negative values triggers the phase transformation. In other words, the ultimate threshold of the fluid as a thermodynamic phase coincides with the threshold of its dynamic stability as a superfluid. Exactly at this point, the roton wave vector equals $2\pi/a_c$.

To evaluate the energy of the perturbed solution, besides the amplitude we also need to calculate the phase gradient. To this aim, we must solve Eq.\,(\ref{b-11}). The terms on the r.h.s. are estimated as:
\be
\nabla(\nabla^2\delta\eta)=\varepsilon k^2{\bf k}\sin({\bf k}\cdot{\bf x}-\omega t)
\label{b-16}
\ee
and
\be
\nabla\int{\rm d}^dy\,\delta\eta({\bf y},t)u({\bf x}-{\bf y})=-\varepsilon{\bf k}\widetilde{u}(k)\sin({\bf k}\cdot{\bf x}-\omega t)\,.
\label{b-17}
\ee
Hence, the solution to (\ref{b-11}) is $\delta{\bf u}=\chi{\bf k}\cos({\bf k}\cdot{\bf x}-\omega t)$ with
\be
\chi=\varepsilon\frac{m}{\hbar}\frac{\omega(k)}{k^2}\,.
\label{b-18}
\ee
We now substitute $\eta=1+\varepsilon\cos({\bf k}\cdot{\bf x}-\omega({\bf k})t)$ and $\nabla\theta=\chi{\bf k}\cos({\bf k}\cdot{\bf x}-\omega({\bf k})t)$ into the energy functional, which in terms of $\eta$ and $\theta$ reads~\cite{Pomeau}:
\ba
{\cal E}[\eta,\theta]&=&\frac{\rho\widetilde{u}(0)}{2}+\frac{\hbar^2}{8mV}\int{\rm d}^dx\left(\frac{\left(\nabla\eta\right)^2}{\eta}+4\eta(\nabla\theta)^2\right)
\nonumber \\
&+&\frac{\rho}{2V}\int{\rm d}^dx\,{\rm d}^dx'\left(\eta({\bf x}')-1\right)u({\bf x}-{\bf x}')\left(\eta({\bf x})-1\right)\,.
\label{b-19}
\ea
Up to ${\cal O}(\varepsilon^3)$ terms, the energy is given by:
{\small
\ba
{\cal E}&=&\frac{\rho\widetilde{u}(0)}{2}+\frac{\hbar^2k^2\varepsilon^2}{8mV}\int{\rm d}^dx\,\sin^2({\bf k}\cdot{\bf x}-\omega({\bf k})t)+\frac{\hbar^2k^2\chi^2}{2mV}\int{\rm d}^dx\,\cos^2({\bf k}\cdot{\bf x}-\omega({\bf k})t)
\nonumber \\
&+&\frac{\rho\varepsilon^2}{2V}\int{\rm d}^dx\,{\rm d}^dx'\,\cos({\bf k}\cdot{\bf x}'-\omega({\bf k})t)\,u({\bf x}-{\bf x}')\cos({\bf k}\cdot{\bf x}-\omega({\bf k})t)\,.
\label{b-20}
\ea
}
Now imagine that the box $V$ is an hypercube of side $L=V^{1/d}$ and assume that {\bf k} takes the discrete values
\be
{\bf k}=\frac{2\pi}{L}(n_1,n_2,\ldots,n_d)\,\,\,\,\,\,{\rm with}\,\,\,\,\,\,n_\alpha=0,\pm 1,\pm 2,\ldots
\label{b-21}
\ee
We first evaluate the kinetic term. For one thing,
\ba
&&\int{\rm d}^dx\,\sin^2({\bf k}\cdot{\bf x}-\omega({\bf k})t)=\frac{1}{2}\int{\rm d}^dx\left[1-\cos(2{\bf k}\cdot{\bf x}-2\omega({\bf k})t)\right]
\nonumber \\
&=&\frac{V}{2}-\frac{1}{4}\int{\rm d}^dx\left[e^{2i({\bf k}\cdot{\bf x}-\omega({\bf k})t)}+{\rm h.c.}\right]=\frac{V}{2}\,.
\label{b-22}
\ea
Similarly,
\be
\int{\rm d}^dx\,\cos^2({\bf k}\cdot{\bf x}-\omega({\bf k})t)=\frac{V}{2}\,.
\label{b-23}
\ee
In the end, the kinetic energy reads:
\ba
&&\frac{\hbar^2k^2\varepsilon^2}{8mV}\int{\rm d}^dx\,\sin^2({\bf k}\cdot{\bf x}-\omega({\bf k})t)+\frac{\hbar^2k^2\chi^2}{2mV}\int{\rm d}^dx\,\cos^2({\bf k}\cdot{\bf x}-\omega({\bf k})t)
\nonumber \\
&=&\frac{\hbar^2k^2}{2m}\frac{\varepsilon^2}{8}+\frac{m\omega^2(k)}{4k^2}\varepsilon^2\,.
\label{b-24}
\ea
As for the potential energy,
\ba
&&\int{\rm d}^dx\,{\rm d}^dx'\,\cos({\bf k}\cdot{\bf x}'-\omega({\bf k})t)\,u({\bf x}-{\bf x}')\cos({\bf k}\cdot{\bf x}-\omega({\bf k})t)
\nonumber \\
&=&\frac{1}{4}\left\{e^{2i\omega({\bf k})t}\int{\rm d}^dx\,{\rm d}^dx'\,e^{-i{\bf k}\cdot({\bf x}+{\bf x}')}u({\bf x}-{\bf x}')+{\rm h.c.}\right\}
\nonumber \\
&+&\frac{1}{4}\left\{\int{\rm d}^dx\,{\rm d}^dx'\,e^{-i{\bf k}\cdot({\bf x}-{\bf x}')}u({\bf x}-{\bf x}')+{\rm h.c.}\right\}
\nonumber \\
&=&\frac{1}{4}\left\{e^{2i\omega({\bf k})t}\widetilde{u}(k)\underbrace{\int{\rm d}^dx'\,e^{-2i{\bf k}\cdot{\bf x}'}}_{0}+{\rm h.c.}\right\}+\frac{V}{2}\widetilde{u}(k)=\frac{V}{2}\widetilde{u}(k)\,.
\label{b-25}
\ea
Plugging Eqs.\,(\ref{b-24}) and (\ref{b-25}) into Eq.\,(\ref{b-20}), the specific energy of the perturbed solution finally equals
\be
{\cal E}=\frac{\rho\widetilde{u}(0)}{2}+\frac{\varepsilon^2}{8}\left(\frac{\hbar^2 k^2}{2m}+2\rho\widetilde{u}(k)\right)+\frac{m\omega^2(k)}{4k^2}\varepsilon^2=\frac{\rho\widetilde{u}(0)}{2}+\frac{m\omega^2(k)}{2k^2}\varepsilon^2\,,
\label{b-26}
\ee
which is clearly larger than the homogeneous-fluid energy.

\section{Superfluid fraction of the crystal}
\renewcommand{\theequation}{C.\arabic{equation}}

Like a superfluid, also a supersolid can be characterized by the nature of its response to uniform axial rotations~\cite{Leggett,Sepulveda}. Under a slow rotation, a fraction of the quantum solid may stand still, with the result that its moment of inertia is smaller than expected from a classical analysis. Leggett~\cite{Leggett} has proposed to call {\em superfluid fraction} of a quantum solid the quantity (also dubbed ``non-classical rotational inertia fraction''):
\be
f_s=\frac{I_0-I}{I_0}\,,
\label{c-1}
\ee
where $I$ is the moment of inertia of the crystal around the axis of rotation and $I_0$ its classical value.

To estimate $f_s$ we appeal to an argument in Ref.~\cite{Aftalion}, which we here adapt to our setting. We first recall that, when a thermodynamic system is subject to rotation, say, around the $z$ coordinate axis, the free energy at $T=0$ and $P=0$ should be written as:
\be
E-\omega L_z\,,
\label{c-2}
\ee
$E$ being the total energy in the presence of rotation, $\omega$ the angular velocity, and $L_z$ the $z$-component of the angular momentum (notice that the first law of thermodynamics in differential form reads: ${\rm d}E=T{\rm d}S-P{\rm d}V+\omega{\rm d}L_z+\mu{\rm d}N$). For a system of rotating bosons, the state $\psi$ must be determined by requiring that the energy functional $e[\psi]-\omega\left\langle\psi|L_z|\psi\right\rangle$ be minimum, which for small $\omega$ values is nothing but the energy per particle in the absence of rotation {\em minus} $(1/2)I\omega^2$. In other words:
\be
I=-\left.\frac{\partial^2}{\partial\omega^2}{\rm min}_{\psi}\left\{e[\psi]-\omega\left\langle\psi|L_z|\psi\right\rangle\right\}\right|_{\omega=0}\,.
\label{c-3}
\ee
The operator $L_z$ is given by
\be
\left\langle\psi|L_z|\psi\right\rangle=-\frac{i\hbar}{2}\hat{\bf z}\cdot\int{\rm d}^dx\,{\bf r}\wedge\left(\psi^*\nabla\psi-\psi\nabla\psi^*\right)\,.
\label{c-4}
\ee
For $\omega\ne 0$ the quantum state $\psi$ acquires a phase, $\theta({\bf x})=\omega S({\bf x})+{\cal O}(\omega^2)$ (like in \cite{Aftalion}, we assume that the amplitude $\sqrt{\eta}$ is instead the same as without rotation; this statement is tantamount to saying that any possible $\omega$-dependence of $\eta$ can only have relevance for the properties of the ``normal'' solid component). Putting $\psi=(1/\sqrt{V})\sqrt{\eta}\exp\{i\omega S\}$ in (\ref{c-4}), we readily obtain:
\be
\left\langle\psi|L_z|\psi\right\rangle=\frac{\hbar\omega}{V}\int{\rm d}^dx\,\eta\nabla S\cdot(\hat{\bf z}\wedge{\bf r})\,,
\label{c-5}
\ee
leading in turn (see (\ref{b-19})) to
\be
e[\psi]-\omega\left\langle\psi|L_z|\psi\right\rangle=e_0[\psi]+\frac{\hbar^2}{2m}\omega^2\frac{1}{V}\int{\rm d}^dx\,\eta(\nabla S)^2-\frac{\hbar\omega^2}{V}\int{\rm d}^dx\,\eta\nabla S\cdot(\hat{\bf z}\wedge{\bf r})\,,
\label{c-6}
\ee
where $e_0[\psi]$ is the energy functional for $\theta=0$. Therefore:
\be
I={\rm min}_S\left\{\frac{2\hbar}{V}\int{\rm d}^dx\,\eta\nabla S\cdot(\hat{\bf z}\wedge{\bf r})-\frac{\hbar^2}{m}\frac{1}{V}\int{\rm d}^dx\,\eta(\nabla S)^2\right\}\,.
\label{c-7}
\ee
Upon considering that
\be
I_0=\frac{1}{V}\int{\rm d}^dx\,\eta mr_\perp^2=\frac{m}{V}\int{\rm d}^dx\,\eta(\hat{\bf z}\wedge{\bf r})^2\,,
\label{c-8}
\ee
we finally obtain:
\be
f_s=\frac{\hbar^2}{m^2}\frac{{\rm min}_S\left\{\int{\rm d}^dx\,\eta\left(\nabla S-(m/\hbar)\hat{\bf z}\wedge{\bf r}\right)^2\right\}}{\int{\rm d}^dx\,\eta(\hat{\bf z}\wedge{\bf r})^2}\,.
\label{c-9}
\ee
While computing $f_s$ is difficult, finding a lower value is much simpler, if we consider that
\be
f_s\ge\frac{\eta_{\rm min}}{\eta_{\rm max}}\frac{\hbar^2}{m^2}\frac{{\rm min}_S\left\{\int{\rm d}^dx\left(\nabla S-(m/\hbar)\hat{\bf z}\wedge{\bf r}\right)^2\right\}}{\int{\rm d}^dx(\hat{\bf z}\wedge{\bf r})^2}=\frac{\eta_{\rm min}}{\eta_{\rm max}}\,.
\label{c-10}
\ee
To obtain this estimate, we have made use of the fact that the minimum of $\int{\rm d}^dx\left(\nabla S-(m/\hbar)\hat{\bf z}\wedge{\bf r}\right)^2$ is reached for $\nabla S=0$ (the argument goes as follows: first note that there is no gradient equal to $(m/\hbar)(\hat{\bf z}\wedge{\bf r})$, since $\nabla\wedge(\hat{\bf z}\wedge{\bf r})\ne 0$; on the other hand, the Euler-Lagrange equation for the functional in (\ref{c-10}) is $\nabla^2S=0$, and the only bounded harmonic function on $\mathbb{R}^3$ is a constant). It is clear that in our variational theory the function $\eta$ never vanishes in the middle region between one lattice site and the other, implying that $f_s$ is strictly positive at every pressure (in other words, the Gaussian variational crystal is {\em supersolid}).

\end{document}